\documentclass[12pt, draftclsnofoot, onecolumn]{IEEEtran}
 \usepackage{amsmath,amssymb}
 \usepackage{subfigure}
 \usepackage{graphicx,graphics,color,psfrag}
 \usepackage{cite,balance}
 \usepackage{caption}
 \allowdisplaybreaks
 \usepackage{algorithm}
 \usepackage{accents}
 \usepackage{amsthm}
 \usepackage{bm}
  \usepackage{url}
 \usepackage{algorithmic}
 \usepackage[english]{babel}
 \usepackage{multirow}
 \usepackage{enumerate}
 \usepackage{cases}
 \usepackage{stfloats}
 \usepackage{dsfont}
 \usepackage{color,soul}
 \usepackage{amsfonts}
 \usepackage{cite,graphicx,amsmath,amssymb}
 \usepackage{subfigure}
 \usepackage{fancyhdr}
 \usepackage{hhline}
 \usepackage{graphicx,graphics}
 \usepackage{array,color}
 \usepackage{amsmath}

\newtheorem{definition}{Definition}

\newtheorem{lemma}{Lemma}
\newtheorem{corollary}{Corollary}

\newtheorem{proposition}{Proposition}

\newtheorem{assumption}{Assumption}

\newtheorem{remark}{\bf Remark}
\def\phi{\varphi}

\def\l{\left}
\def\r{\right}
\def\({\left(}
\def\){\right)}

\def\b0{{\mathbf{0}}}

\newcommand{\nn}{\nonumber}

\begin{document}
\title{\huge   {\color{black}{Exploiting Non-Causal CPU-State Information for Energy-Efficient Mobile Cooperative Computing}} } 
\author{Changsheng You and  Kaibin Huang  \thanks{\noindent C. You and K. Huang are  with the Dept. of EEE at The  University of  Hong Kong, Hong Kong (Email: csyou@eee.hku.hk, huangkb@eee.hku.hk).}}
\maketitle

\vspace{-18pt}
\begin{abstract} Scavenging the idling computation resources at the enormous number of mobile devices, ranging from small IoT devices to powerful laptop computers, can provide a powerful platform for local mobile cloud computing. The vision can be realized by peer-to-peer cooperative computing between edge devices, referred to as \emph{co-computing}. {\color{black}{This paper exploits non-causal helper's CPU-state information to design energy-efficient co-computing policies for scavenging time-varying spare computation resources at peer mobiles.}}  Specifically, we consider a co-computing system where a user offloads computation of input-data to a helper. The helper controls the offloading process for the objective of minimizing the user's energy consumption based on a predicted helper's CPU-idling profile that specifies the amount of  available computation resource for co-computing. Consider the scenario that the user has one-shot input-data arrival and the helper buffers offloaded bits. 
{\color{black}{The problem for energy-efficient co-computing is formulated as two sub-problems: the slave problem corresponding to adaptive offloading and the master one to data partitioning. Given a fixed offloaded data size, the adaptive offloading aims at minimizing the energy consumption for offloading by controlling the offloading rate under the deadline and buffer constraints. By deriving the necessary and sufficient conditions for the optimal solution, we characterize the structure of the optimal policies  and propose algorithms for computing the policies.}} 
%The derived solution for the optimal offloading control has an interesting graphical interpretation as follows. In the plane of user's co-computable bits (by offloading) versus time, a so-called \emph{offloading feasibility tunnel} can be constructed that constrains the range of offloaded bits at any time instant. The existence of the tunnel arises from the helper's CPU-idling profile and buffer size. Given the tunnel, the optimal offloading is shown to be achieved by the well-known ``\emph{string-pulling}" strategy, graphically referring to pulling a string across the tunnel.
 Furthermore, we show that the problem of optimal data partitioning for offloading and local computing at the user is convex, admitting a simple solution using the sub-gradient method. {\color{black}{Last, the developed design approach for co-computing is extended to the scenario of bursty data arrivals at the user accounting for data causality constraints. Simulation results verify the effectiveness of the  proposed algorithms.}}
% The approach is modified by  defining a new offloading feasibility tunnel that accounts for bursty  data arrivals.
\end{abstract}
%\begin{IEEEkeywords}
%Mobile cooperative computing, energy-efficient transmission, D2D communication, computation offloading, mobile-edge computing.
%\end{IEEEkeywords}
%\newpage
\section{Introduction}There exist tens of billions of mobile devices distributed at network edges such as smartphones and laptop computers. They are equipped with powerful CPUs but a large population are idle at any given time instant. Scavenging the enormous amount of  distributed computation resources can provide a new platform for mobile cloud computing and furthermore alleviate the problems of network congestion and long latency for the classic cloud computing. This vision has been driving extensive research 
%and development
 in both the academia and industry under various names such as \emph{edge computing} and \emph{fog computing}{\color{black}{\cite{taleb2017multi,mao2017mobile2,taleb2017mobile,chiangfog}}}. One technology for materializing the vision is \emph{mobile cooperative computing}, namely the cooperation between mobiles in computing by sharing computation resources and thereby improving their utilizations. This technology, referred to as \emph{co-computing} for simplicity, is the theme of this paper. {\color{black}{Specifically, this paper presents co-computing algorithms for enabling energy-efficient \emph{peer-to-peer} (P2P) computation offloading that \emph{exploits CPU-state information for scavenging spare computation  resources at mobiles.
}}}
%  \emph{opportunistically exploits time-varying spare computation resources at mobiles}. 
\subsection{\color{black}{Related Work}}
\subsubsection{\color{black}{Multi-Access Edge Computation Offloading}}
\emph{Mobile edge computing} (MEC), initiated by ETSI, refers to providing mobiles with cloud-computing capabilities  and IT service from \emph{base stations} (BSs) or \emph{access points} (APs) at the edge of mobile networks. {\color{black}{It was renamed as multi-access edge computing as its applications have been broadened to include radio access networks (including WiFi) and multiple-access technologies \cite{taleb2017multi}.}}  The recent inclusion of MEC on the roadmap of developing next-generation network architecture has motivated active research on developing wireless techniques for offloading \cite{patel2014mobile}.
%\cite{ETSI_Framework2}. 
{\color{black}{This has led to the emergence of an active area, called \emph{multi-access edge computation offloading} (MECO), that merges  two  disciplines: wireless communications and mobile computing \cite{mao2017mobile2}}}. Making a binary decision on offloading-or-not involves a straightforward comparison of mobile-energy  consumption for computing given data by offloading and local computing. {\color{black}{However, compared with  traditional traffic offloading \cite{chen2015energy} and green wireless-communication design \cite{wu2012green}, designing   computation  offloading  is more challenging  as it has to jointly  consider two different objectives,  energy-efficient computing and energy-efficient transmissions, in a more complex  system for MECO.}} In particular, energy-efficient techniques are designed in \cite{zhang:MobileMmodel:2013} for controlling the CPU frequency for local computing and transmission rate for offloading. They are integrated with wireless energy transfer technology in \cite{you2015energyJSAC} to power mobiles for enhancing mobile energy savings. By program partitioning, a task can be divided for  \emph{partial offloading} (and partial local computing) \cite{mao2017mobile2}. Various approaches have been developed for partial offloading  such as live (in-computing) prefetching of mobile data to the server for reducing communication overhead \cite{ko2017live} and optimal program partitioning using integer programming \cite{MahmoodiTCC16}. 

The design of multiuser MECO systems involves the new research issue of joint radio-and-computation resource allocation\cite{you2016energy,chen2015efficient,lyumulti:2016:ProxiCloud,mao2017stochastic} for achieving system-level objectives (e.g., minimum sum mobile-energy consumption). Specifically, the \emph{centralized} resource allocation is studied in \cite{you2016energy}, where an offloading priority function is derived to facilitate making binary offloading decisions for individual users. On the other hand, algorithms for  \emph{distributed} resource allocation are designed in \cite{chen2015efficient}, \cite{lyumulti:2016:ProxiCloud} by solving formulated integer optimization problems using game theory and decomposition techniques.  Last, server scheduling is also a relevant topic for designing multiuser MECO systems and has been studied in \cite{guo2016energy,yang2015multi,zhao2015cooperative} for coping with various  issues including heterogeneous latency requirements, sub-task dependency and cloud selection, respectively.

MEC and MECO are enabled by the edge clouds implemented by dedicated servers (e.g., BSs or APs). However, in view of the exponentially-increasing IoT devices and computation traffic, the massive users accessing the servers will incur overwhelming communication overhead and soon exhaust the servers' capacities. On the other hand, latest mobile devices, e.g., smartphones and laptop computers equipped with multi-core processors, are comparable with normal servers in terms of computing power. Scavenging the excessive computation resources in massive idling mobile devices drives active research on co-computing discussed in the sequel.
%next sub-section. 
\subsubsection{Mobile Cooperative Computing}\label{Sec:IntroCo-Computing}
Recent research on mobile co-computing is characterized by the themes of resource sharing and cooperative computing \cite{song2014energy,wang2016cooperative,pu2016d2d,XCao1704,sheng2015energy}. An online algorithm is proposed in \cite{song2014energy} for implementing co-computing and result-sharing, and thereby achieving the optimal energy-and-traffic tradeoff. Since users have no commitments for cooperation, one important aspect of co-computing research is to design schemes for incentivizing them for sharing computation resources, where a suitable design tool  is game theory adopted in \cite{wang2016cooperative}. From the aspect of wireless communication, P2P offloading in co-computing can be efficiently implemented using the well developed \emph{device-to-device} (D2D)  communication technology. This direction is pursued in \cite{pu2016d2d} where offloading based on D2D transmissions is controlled by a cooperation policy optimized using Lyapunov optimization theory. {\color{black}{In addition,  let a \emph{helper} refer to a cooperative mobile that shares computation resources with peers. A  joint computation-and-communication cooperation protocol is proposed in \cite{XCao1704}, where the helper not only computes part of the tasks offloaded from the user, but also acts as a relay node to forward the tasks to the MEC server.}} Last, an interesting type of sensor networks is proposed in \cite{sheng2015energy} to implement co-computing between sensors based on the discussed partial offloading technique.

In view of the above prior work, one key fact that is overlooked is that  the \emph{non-causal CPU-state information} (NC-CSI)\footnote{\color{black}{Causal information refers to information on present and past events, while non-causal information is on future events.}} referring to the time profile of CPU state, can be exploited by the helper to design energy-efficient co-computing.
%control the offloading rates of a user, so as to simultaneously optimize the resource utilization and reduce the transmission-energy consumption. }}
Acquiring such information is feasible by leveraging advancements in two areas, namely \emph{CPU profiling} and \emph{CPU-utilization prediction}. The former measures the usage of computation tasks by constructing CPU profile trees \cite{chun2011clonecloud} or integrating the CPU distribution and time-series profiles \cite{wood2007black}. In the industry, CPU profiling has been implemented by e.g., Apple Inc.,  via tracking the core-and-thread usage by devices.
%\footnote{https://developer.apple.com/library/content/documentation/DeveloperTools/Conceptual/InstrumentsUserGuide/MeasuringCPUUse.html}  
{\color{black}{On the other hand, leveraging the time correlation of computation loads,  the short-term CPU utilization (e.g., a few seconds) can be predicted based on simple linear models, such as \emph{autoregressive moving average} (ARMA) model in \cite{dinda1999evaluation}. While the long-term CPU utilization can be modeled by a non-linear function; its prediction requires sophisticated techniques from machine learning, such as Bayesian learning and \emph{Gaussian process regression} (GPR) \cite{bui2016energy} which is non-parametric without specifying the prediction parameters. More details about the prediction-model selection can be found in \cite{islam2012empirical}.}}   The availability of technologies for CPU profiling and utilization prediction motivates the current design to exploit NC-CSI for improving the performance of co-computing.

%Thereby, this work proposes a novel concept that computation prediction can also help energy-efficient transmissions in the context of mobile edge computing. 

%  two design challenges tackled in this work. First, energy-efficient P2P computation offloading must be adaptive to the time-varying cascaded channels. Second, to avoid overloading helper's CPU, ``transmission" over the computation channel should rely on the helper-CPU resource whose usage must satisfy the \emph{real-time constraints}. Specifically, CPU cycles available at a particular time instant must be used in \emph{real-time} but not earlier or later.}} In contrast, stored energy for transmission over the wireless channels allows flexible usage in time.
\begin{figure}[t!]
\begin{center}
\includegraphics[width=11.5cm]{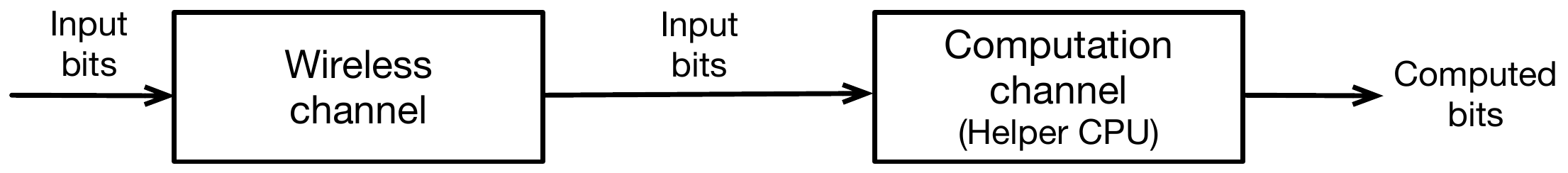}
%\vspace{-2pt}
\caption{Cascaded wireless-and-computation channels for mobile co-computing.}
\label{Fig:Channels}
\end{center}
\end{figure}
\vspace{-3pt}
\vspace{-3pt}
\subsection{{\color{black}{Motivations and Contributions}}}
{\color{black}{In this work, leveraging the advantages of NC-CSI, we contribute to the area of mobile co-computing by addressing two new issues. \emph{The first is how to exploit NC-CSI for opportunistically scavenging spare computation resources.} One key characteristic of co-computing is that a helper assigns a higher priority for computing local tasks and their random arrivals result in time variations in the computation resources available for sharing.  The existing designs for co-computing are unable to fully scavenge dynamic computation resources at a helper due to transmission latency. In the current work, we propose a novel solution for overcoming this drawback by exploiting NC-CSI acquired from computation prediction. This allows a mobile to plan transmission \emph{in advance} so as to fully utilize random CPU idling periods at a helper. 
%Thereby, this work shows the significance of computation prediction for mobile cooperative computing, pointing to many potential opportunities in emerging areas such as Fog computing.

The second issue not yet address in prior work is \emph{how to exploit NC-CSI for minimizing mobile energy consumption.} Note that the said dynamic spare resources create a virtual \emph{computation channel} where the channel throughput is the number of computed bits. This gives an interesting interpretation of co-computing as adaptive transmission over the cascaded \emph{wireless-and-computation channels} shown in Fig.~\ref{Fig:Channels}. Such interpretation gives rise to the following design challenges for minimizing mobile energy consumption. On one hand, transmitting offloaded data from a mobile to a helper too far advance before the helper's CPU is available will necessarily increase the data rate and thus mobile energy consumption. On the other hand, transmitting data too late will miss the opportunities of using the helper's CPU. In other words, ``transmission" over the computation channel should rely on the helper-CPU resource whose usage must satisfy the \emph{real-time constraints}. Specifically, CPU cycles available at a particular time instant must be used in \emph{real-time} but not earlier or later. This is contrast to stored energy for transmission over wireless channels that allows flexible usage in time.  The above dilemma is solved in this work by exploiting NC-CSI to minimize mobile transmission-energy consumption while fully utilize the helper's random computation resource.
 }} 
 
{\color{black}{To the best of the authors' knowledge, this work presents the first attempt to exploit NC-CSI for scavenging spare computation resources at the helper and minimizing mobile energy consumption for mobile co-computing systems. The specific system model and contributions are summarized as follows.}} 
 
Consider a mobile co-computing system comprising one helper and one user, both  equipped with single antenna. The user needs to process the input data for a particular computation task before a given deadline. The input data arrives at the user either at a single time instant or spread over the time duration before the deadline, referred to as \emph{one-shot} and \emph{bursty} arrivals, respectively. Based on the model of partial offloading, the user splits the input data for processing locally and at the helper, leading  to the problem of \emph{data partitioning}.  Consider the mobile user. To model the energy consumption in local computing, it is assumed that processing a single bit requires a fixed number of CPU cycles, each of which consumes a fixed amount of energy. Moreover, the transmission-energy consumption incurred in the offloading process depends on the rate based on the Shannon's equation. Next, consider the helper. The available computation resource for co-computing is modeled as a fixed monotone-increasing curve in the plane of computable bits versus time, called the \emph{helper's CPU-idling profile}. Assume that the helper uses a buffer to store data transmitted by the user and has non-causal knowledge of the profile as well as other information including the channel and local computing energy. Using this information, it controls the transmission by the user, leading to the problem of \emph{adaptive offloading}.

The main contributions of the work  are summarized as follows.
\begin{enumerate}
{\color{black}{
\item{\emph{Adaptive Offloading with One-Shot Data Arrival:} Consider one-shot data arrival at the user. Given a fixed number of input-data bits for offloading and co-computing, the said problem of adaptive offloading is formulated to minimize the transmission-energy consumption under the deadline and  buffer constraints. This complex problem is solved as follows. First, for the large buffer case where the buffer size at the helper is no smaller than the offloaded bits, the formulated non-convex problem is equivalently transformed into a convex problem. By deriving the necessary and sufficient conditions for the optimal solution, we characterize the structure of the optimal policy  and present algorithms for computing it. Geometrically, the optimal policy involves  finding a shortest path under the constraints of the  helper's CPU-idling profile and buffer size. On the other hand, the corresponding problem for the smaller buffer case is still non-convex. To tackle this challenge, we propose a tractable  approach called \emph{proportional CPU-utilization} and prove that it is asymptotically optimal.}
\item{\emph{Energy-Efficient Data Partitioning with One-Shot Data Arrival:} Next, building on the solution for adaptive offloading, the said data partitioning problem is formulated to minimize user's energy consumption. Directing solving this problem is intractable due to the lack of closed-form expression for the objective function. We address this difficulty by proving that the formulated problem is convex even without a closed-form expression. Then a sub-gradient method is applied to compute the optimal data-partitioning policy.  }
\item{\emph{Mobile Co-Computing with Bursty Data Arrivals:} The versatility of the above solution approach is  further demonstrated by an extension to the case with bursty data arrivals at the user. For tractability, we consider a simple scheme of \emph{proportional data partitioning} for each instant of data arrival using a uniform ratio, which is an optimization variable. Accounting for the data causality constraints, i.e., the input-data bit cannot be offloaded or computed before it arrives, the corresponding adaptive offloading and data partitioning policies can be modified from the counterparts with one-shot data arrival.}} 
 }
\end{enumerate}

%The reminder of this paper is organized as follows. Section~\ref{Sec:Sys} introduces the system model. In Section~\ref{Sec:EgyOneShot}, energy-efficient co-computing policies are derived for the case of one-shot data arrival at the user and a large buffer at the helper. 
%The corresponding solution approach is extended to account for the cases of a smaller buffer at the helper in Section~\ref{Sec:SmallBuffer} and bursty data arrivals at the user in Section~\ref{Sec:RandomComputLoad}, respectively. Last, simulation results and discussion are given in Section~
%\ref{Sec:Simu}.
%, followed by the conclusion in Section~VII. 
%\vspace{-5pt}
\section{System Model}\label{Sec:Sys}
\begin{figure}[t!]
\begin{center}
\includegraphics[width=14cm]{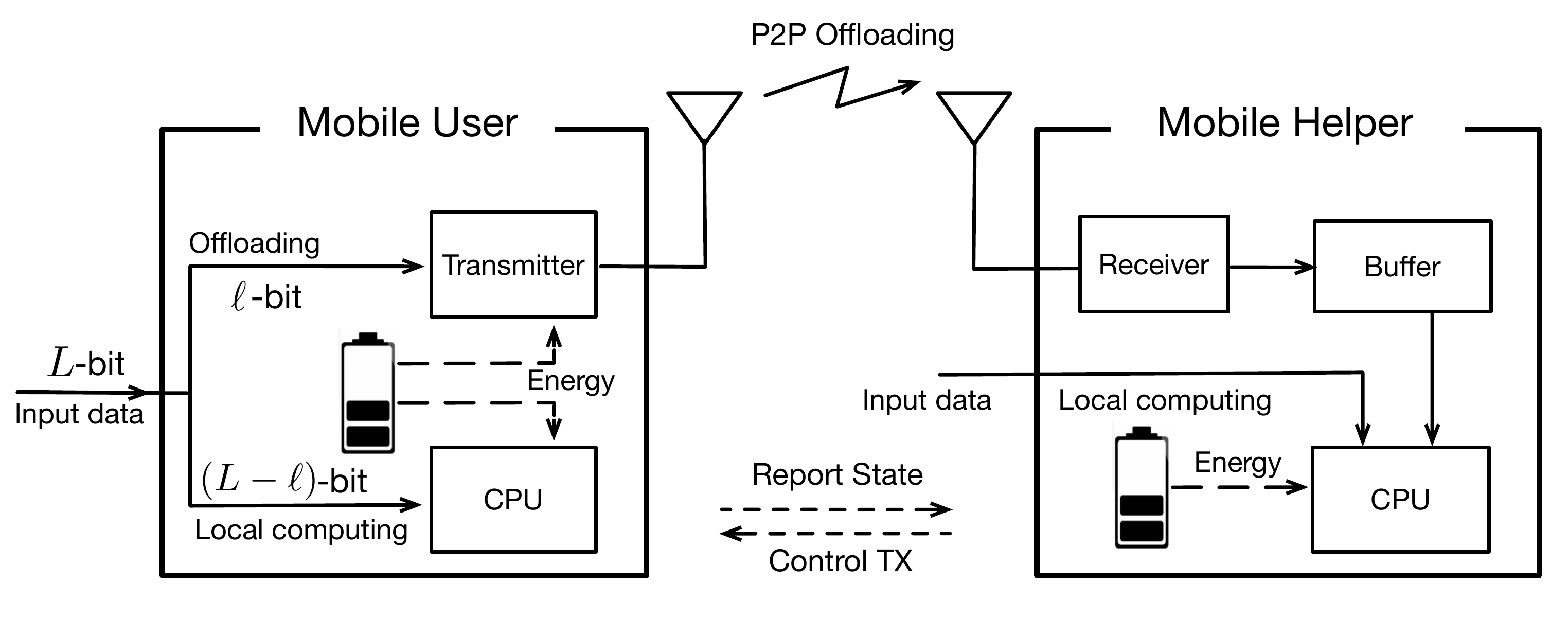}
\vspace{-1pt}
\caption{Model of one co-computing system.}
\label{Fig:SysMobile}
\end{center}
\end{figure}

Consider one co-computing system shown in Fig.~\ref{Fig:SysMobile}, comprising one user and one helper both equipped with single antenna\footnote{\color{black}{For simplicity, we consider a system comprising one helper serving one user. However, the random CPU state at the helper  implies that the helper serves another user or locally generated tasks in the CPU-busy time (see Fig.~\ref{Fig:PrimaryCPUState}). Then the user in the current system aims at scavenging the remaining random CPU-idling time at the helper.}}. The user is required to finish a computation task with either one-shot or bursty input-data arrival before a deadline $T$. To this end, it adaptively offloads partial/all data to the helper for co-computing based on the control policy developed at the helper. The helper operates at a constant CPU frequency but with intermittent local computing tasks. It is assumed that the helper has sufficient energy for receiving and computing the data from the user\footnote{{\color{black}{Before offloading, the user is assumed to send a probing signal to the helper and receive feedback comprising NC-CSI as well as information on whether the helper has sufficient energy for cooperation.}}  }.  The specific models and assumptions are described in the following sub-sections.

%\vspace{-5pt}
\subsection{Model of the Helper's CPU-Idling Profile}
 The \emph{helper's CPU-idling profile} is defined as the user's data (in bits) that can be computed by the helper in the duration $t\in[0,T]$, which is denoted as $U_{{\rm{bit}}}(t)$ and modeled shortly.
% \vspace{-3pt}
\begin{definition}[Helper-CPU State Information]\label{Def:Primary}\emph{Helper-CPU state information refers to the CPU state over time, which can be modeled by the helper-CPU event space, process and epochs defined as follows. Let $\boldsymbol{\mathcal{E}}=\{\mathcal{E}_1, \mathcal{E}_2\}$ denote the helper-CPU's \emph{event space}, where $\mathcal{E}_1$ and $\mathcal{E}_2$ denote the events that the helper-CPU  changes the state from busy-to-idle and from idle-to-busy, respectively. The helper-CPU  \emph{process} can be then defined as the time instants for a sequence of helper-CPU  events $\{\mathcal{E}_2, \mathcal{E}_1, \mathcal{E}_2, \cdots\}$: $0=s_0<s_1<s_2<\cdots<s_{\tilde{K}-1}<s_{\tilde{K}}=T$. The time interval between two consecutive events\footnote{{\color{black}{In this work, the events correspond to instantaneous CPU-state transitions and thus the time spent on each event is zero.}} } is called an \emph{epoch} with length $\tau_k=s_{k}-s_{k-1}$, for $k=1,\cdots, \tilde{K}$}.
\end{definition} 
\begin{figure}[t!]
\centering
\subfigure[Helper-CPU's event space, process  and epochs.]{\label{Fig:PrimaryCPUState}
\includegraphics[width=11cm]{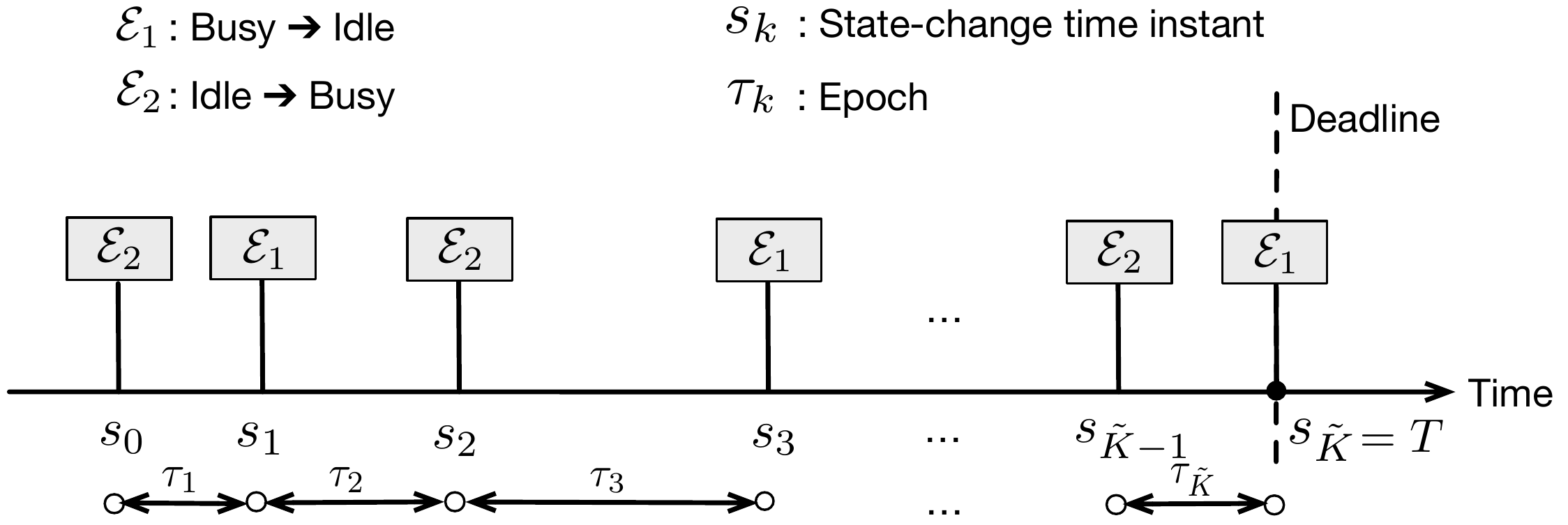}}
\subfigure[Helper's CPU-idling profile.]{\label{Fig:PrimaryCPUProfile}
\includegraphics[width=10.5cm]{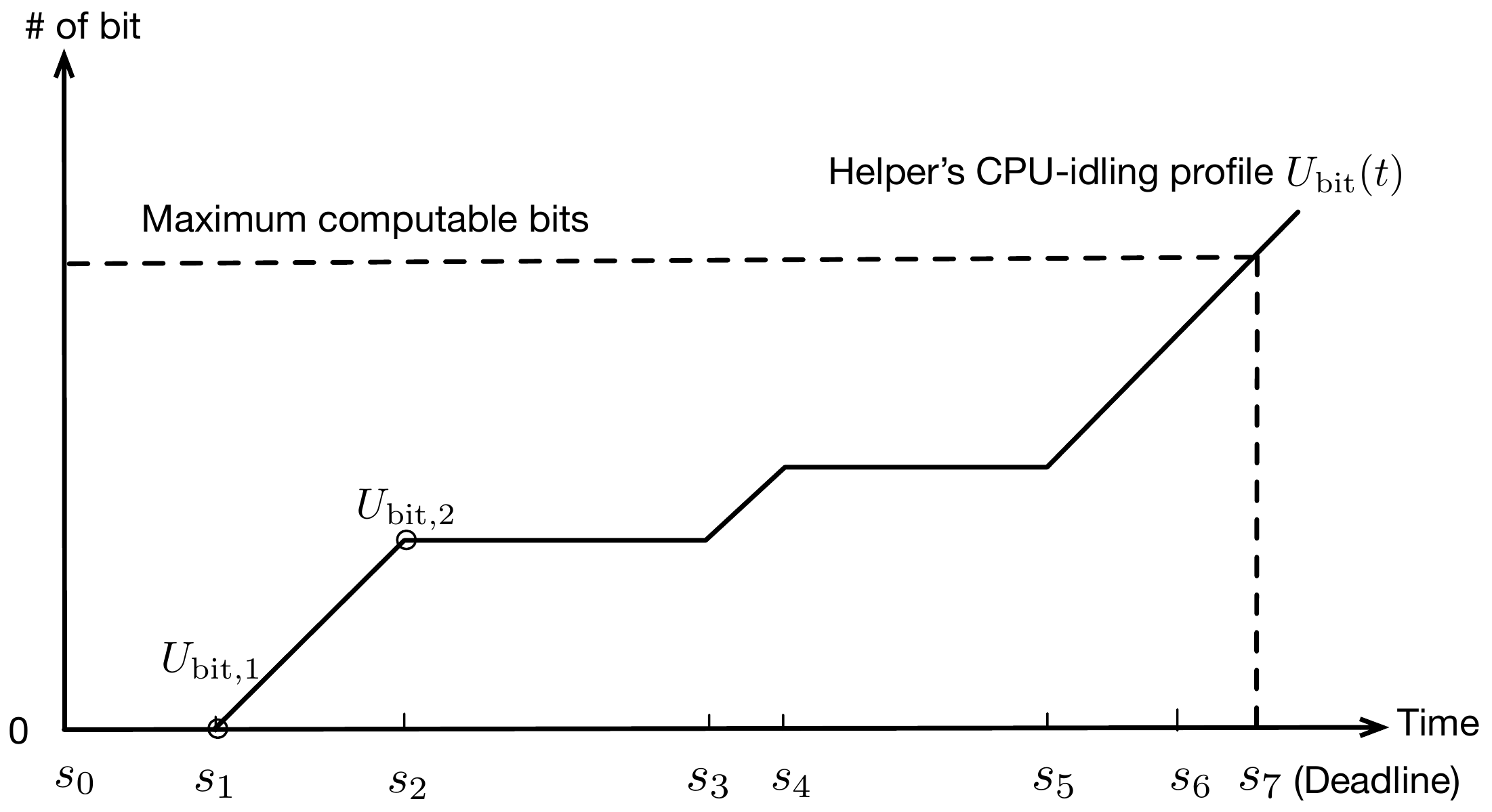}}
\caption{Model of  the helper's CPU process and CPU-idling profile.}
\end{figure}
\vspace{-5pt}
\begin{assumption}\label{Ass:NC-CSI}\emph{The helper has \emph{non-causal helper-CPU state information.} }
\end{assumption}
{\color{black}{This assumption corresponds to the case where the helper performs CPU profiling or predicts the CPU utilization by e.g., linear regression  \cite{dinda1999evaluation} or machine learning \cite{bui2016energy} (see discussion in Section~\ref{Sec:IntroCo-Computing}).}} It allows the off-line design of co-computing policies in the sequel. 

 One sample path of the helper-CPU's process is shown in Fig.~\ref{Fig:PrimaryCPUState}.
For each epoch, say epoch $k$, let $a_k$ represent the CPU-state indicator, where the values of 1 and 0 for $a_k$ indicate the idle and busy states, respectively. Moreover, let $f_h$ denote the constant CPU frequency of the helper and $C$ the number of CPU  cycles required for computing $1$-bit of input-data of the user. Based on the above definitions, the helper's CPU-idling profile can be modeled as
\begin{equation}\label{Eq:CPUIdlingProfile}
U_{{\rm{bit}}}(t)=\l[\sum_{k=1}^{\bar{k}(t)} a_k \tau_k+ a_{\bar{k}(t)+1} \l(t-\sum_{k=1}^{\bar{k}(t)} \tau_k\r)\r] \frac{f_h}{C}, \quad 0\le t\le T,
\end{equation}
where $\bar{k}(t)=\max\{k: \sum_{j=1}^{k} \tau_j \le t\}$, as illustrated in Fig.~\ref{Fig:PrimaryCPUProfile}.  Observe from the figure that the profile can be also represented by a sequence $\{U_{{\rm{bit}},1}, U_{{\rm{bit}},2}, \cdots\}$, with $U_{{\rm{bit}},k}=U_{{\rm{bit}}}(s_k)$. Based on Assumption~\ref{Ass:NC-CSI}, the helper has non-causal knowledge of helper's CPU-idling profile. Last, the helper is assumed to reserve a $Q$-bit buffer for storing the offloaded data before processing in the CPU as shown in Fig.~\ref{Fig:SysMobile}.

\subsection{Models of Local Computing and Offloading}\label{Mod:SU}
\begin{figure}[t]
\begin{center}
\includegraphics[width=11cm]{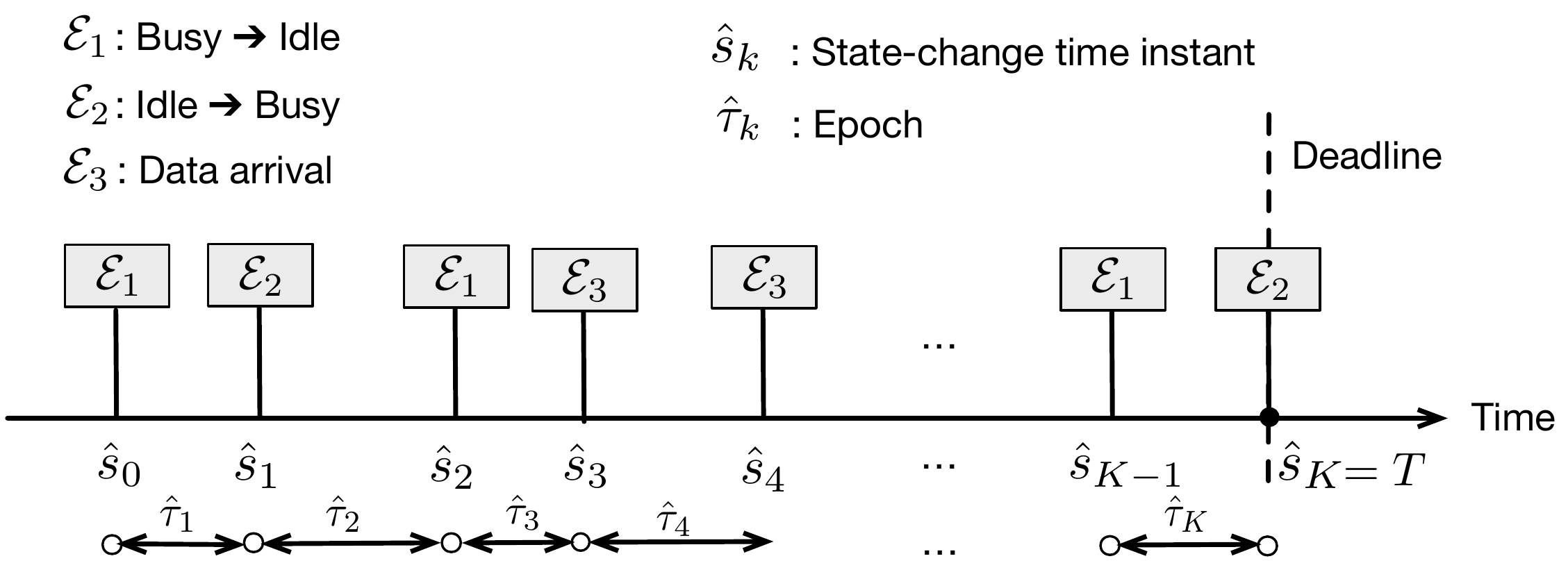}
%\vspace{-2pt}
\caption{Combined stochastic process for bursty data arrivals.}
\label{Fig:SecPri}
\end{center}
\end{figure}

Consider both forms of data arrivals at the user. The one-shot data arrival assumes that an $L$-bit input data arrives at time $t=0$ and thus the helper-CPU's event space and process follow from Definition~\ref{Def:Primary}. On the other hand, the bursty data arrivals form a stochastic process. For ease of exposition, it is useful to define a stochastic process combing the two precesses for data arrivals and helper-CPU. The definition is in Definition~\ref{Def:Combined} and illustrated in Fig.~\ref{Fig:SecPri}.

\begin{definition}[Combined Stochastic Process for Bursty Data Arrivals] \label{Def:Combined}\emph{For the case of bursty data arrivals, let $\hat{\boldsymbol{\mathcal{E}}}=\{\mathcal{E}_1, \mathcal{E}_2, \mathcal{E}_3\}$ denote the combined event space where $\mathcal{E}_1$, $\mathcal{E}_2$ are given in Definition~\ref{Def:Primary} and  $\mathcal{E}_3$ denotes the event that new data arrives at the user. The corresponding process is a sequence of variables: $0=\hat{s}_0<\hat{s}_1<\hat{s}_2<\cdots<\hat{s}_{\tilde{K}-1}<\hat{s}_{\tilde{K}}=T$, denoting the time instants for a sequence of events $\{\mathcal{E}_1, \mathcal{E}_2, \mathcal{E}_3, \cdots\}$. Moreover, for each time instant, say $\hat{s}_k$, let $L_k$ denote the size of data arrival where $L_k=0$ for events $\mathcal{E}_1$ and $\mathcal{E}_2$ and $L_k\neq 0$ for event $\mathcal{E}_3$. In addition, $L_{\tilde{K}}=0$, otherwise the data arriving at the deadline cannot be computed. Then the total input data $L=\sum_{k=1}^{\tilde{K}} L_k.$ }
\end{definition}
\begin{assumption}\emph{The user has \emph{non-causal} knowledge of  bursty data arrivals in the duration $[0, T]$.}
\end{assumption}
{\color{black}{ The assumption (of non-causal knowledge) means that at time $t=0$, the user has the information of future data arrivals in the duration $[0, T]$ including their arrival-time instants and amounts of computation loads. The information can be acquired by computation  prediction techniques  similar to those for CPU-utilization prediction (see discussion in Section~\ref{Sec:IntroCo-Computing}).}} Moreover, the user is assumed to send the information to the helper together with parametric information including  the channel gain, CPU frequency as well as  energy consumption per bit, allowing  the helper to control the operations of offloading and data partitioning. This spares the user from co-computing control that consumes energy.

Based on the definitions and assumptions, the models of local computing and offloading are described as follows. First, consider local computing.  Let $f$ denote the constant CPU frequency at the user. For the case of one-shot data arrival, as shown in  Fig.~\ref{Fig:SysMobile}, the user offloads $\ell$-bit data to the helper and computes the remaining $(L-\ell)$-bit using its local CPU. Due to the deadline constraint, local computing should satisfy: $C(L-\ell)/f\le T.$ It follows that the user  should offload at least  $\ell_{\min}^+$-bit data, where $\ell_{\min}\!=\!L-(f T/C)$ and $(x)^+\!=\!\max\{x, 0\}$.
Following the practical model in \cite{Robert:CMOS:1992}, each CPU cycle consumes the energy of $P_{\rm{cyc}}\!=\!\gamma f^2$ where $\gamma$ is a constant determined by the switched capacitance. As such, $(L-\ell) C P_{\rm{cyc}}$ gives energy consumption for local computing at the user. This model is extended to the case of bursty data arrivals in Section~\ref{Sec:RandomComputLoad}.

Next, consider offloading. For both the cases of one-shot and bursty data arrivals, let $\ell_k$ with $1\le k\le \tilde{K}$ denote the offloaded data size in epoch $k$. 
%$\boldsymbol{\ell}=[\ell_1, \ell_2, \cdots, \ell_{\tilde{K}}]$ denote  the vector of transmitted data bits  where $\ell_k$ for $1\le k\le \tilde{K}$ is the offloaded data size in epoch $k$. 
Since constant-rate transmission within each epoch is energy-efficient \cite{PrabBiyi:EenergyEfficientTXLazyScheduling:2001}, the offloading rate in epoch $k$, denoted by $r_k$, is fixed as $r_k=\ell_k/\tau_k$.
Let $p_k$ represent the transmission power in epoch $k$, then the achievable transmission rate $r_k$ (in bits/s) is $r_k=W \log_{2} \l(1+\dfrac{p_k h^2}{N_0}\r)$
where $h$ is the channel gain and assumed to be fixed throughout the computing duration, $W$ the bandwidth, and $N_0$ the variance of complex-white-Gaussian-channel noise\footnote{\color{black}{In this paper, the D2D interference for co-computing is treated as channel  noise. It is possible for the  helper  to mitigate   the inference by using interference-cancellation techniques, thereby increasing the transmission date rate. However, the proposed design remains largely unchanged except for modifying the noise variance accordingly.}}. Thus, the energy consumption of the user  for offloading $\ell_k$-bit data in epoch  $k$, denoted by $E_k(\ell_k)$, is given as $E_k(\ell_k)=p_k \tau_k=\dfrac{\tau_k}{h^2} f\!\l(\ell_k/\tau_k\r),$
where the function $f(x)$ is defined by $f(x)=N_0 (2^{\frac{x}{W}}-1)$ based on the Shannon's equation.

For ease of exposition, the energy and time the user spends on receiving co-computing results are assumed negligible, as they are typically much smaller than the offloading counterparts. Extending the current analysis to include such overhead is straightforward  though tedious.
\vspace{-5pt}
\subsection{Model of Co-Computing}\label{Sec:MoCocomputing}
 The offloaded data is assumed to be firstly stored in the helper's buffer  and then fetched to the CPU  for co-computing. {\color{black}{To avoid overloading the helper's CPU, we assume that co-computing can be performed only during helper-CPU idle epochs.}} As such, let $T_{\rm{end}}$ and $K =  k(T_{\rm{end}})$ denote the \emph{actual} completion time and corresponding epoch index with $T_{\rm{end}}\le T$ and $K= \tilde{K}$ (or  $\tilde{K}-1$) depending on whether the last epoch is idle (or busy). Note that the idling CPU resource can only be utilized in \emph{real-time} which means \emph{a CPU cycle available now cannot be used in the future}, referred to as the \emph{CPU real-time constraints} in the sequel. Let $d_k(\ell_k)$ denote the computed data size at the helper's CPU during epoch $k$ and $B_k$ the remaining data size at the end of epoch $k$ (or the beginning of epoch $k+1$) with $B_0=0$. Under the CPU real-time constraints, $d_k(\ell_k)$ and $B_k$ evolve as 
 \vspace{-5pt}
\begin{align}\label{Eq:BEvolve}
& \text{(CPU real-time constraints)}\nn\\
% \vspace{-6pt}
& d_k(\ell_k)= \min\l\{ B_{k-1}+\ell_k, \frac{a_k \tau_k  f_h}{C}\r\}, ~~B_k=\sum_{j=1}^k \ell_j-\sum_{j=1}^{k} d_j(\ell_j),  ~~k=1,\cdots, K,
\end{align}
where $(B_{k-1}+\ell_k)$ is the computable data size in epoch $k$ and $(a_k \tau_k f_h/C)$ the available CPU resource. As a result of above constraints, a feasible co-computing strategy should satisfy the following  deadline and buffer constraints.
\begin{itemize}
\item[1)]{\emph{Deadline constraint}: It requires the offloaded $\ell$-bit data to be computed within the deadline:  
\vspace{-10pt}
\begin{equation}\label{Eq:Deadline}
\sum_{k=1}^{K} d_k(\ell_k)=\sum_{k=1}^{K} \ell_k=\ell. 
\end{equation}}
\item[2)]{\emph{Buffer constraints}: Buffer overflow is prohibited, imposing the constraints:
% below at all state-change time instants: 
\vspace{-3pt}
 \begin{equation}\label{Eq:Buffer}
 B_k=\sum_{j=1}^k \ell_j-\sum_{j=1}^{k} d_j(\ell_j)\le Q, ~~ k=1,\cdots, K.
\end{equation}
}
\end{itemize}

\vspace{-5pt}
{\color{black}{
\section{Mobile Cooperative Computing with One-Shot Data Arrival}\label{Sec:EgyOneShot}
In this section, assume that the user has one-shot data arrival and the helper has a finite buffer. We design energy-efficient co-computing algorithms for adaptive offloading and data partitioning. 
\subsection{Problem Formulation}\label{Sec:EGProblem}
Consider that the user has an $L$-bit input-data arrival at time $t=0$. The  problem of energy-efficient co-computing is formulated as  two  sub-problems: the slave problem corresponding to adaptive offloading and the master one to data partitioning.
\subsubsection{Slave Problem of Adaptive Offloading} Given user's  $\ell$-bit offloaded data to the helper, the slave problem aims at minimizing the user's transmission-energy consumption under the deadline and buffer constraints, which can be formulated as:
\begin{equation*}~~({\bf P1})~
\begin{aligned}
\min_ {\boldsymbol{\ell}\ge \boldsymbol{0}}   \quad& \sum_{k=1}^{K} \dfrac{\tau_k}{h^2} f\!\l(\dfrac{\ell_k}{\tau_k}\r) \\
\rm{s.t.}\quad & \sum_{k=1}^{K} d_k(\ell_k)=\sum_{k=1}^{K} \ell_k=\ell, 
\\ & \sum_{j=1}^k \ell_j-\sum_{j=1}^{k} d_j(\ell_j) \le Q, \quad k=1,\cdots, K,
\end{aligned}
\end{equation*}
where $\boldsymbol{\ell}\overset{\triangle}{=}[\ell_1, \ell_2, \cdots, \ell_K]$ and $\boldsymbol{\ell} \ge \boldsymbol{0}$ means that $\ell_k\ge 0, \forall k$.
Let $\{\ell_k^*\}$ solve Problem P1 and thus specify the optimal offloading strategy. Then $E_{\rm{off}}(\ell)=\sum\nolimits_{k=1}^{K} \dfrac{\tau_k}{h^2} f\!\l(\ell_k^*/\tau_k\r)$ denote the minimum transmission-energy consumption.

\subsubsection{Master Problem of Data Partitioning} Given $E_{\rm{off}}(\ell)$, the master problem partitions the $L$-bit data for local computing and offloading. Under the criterion of minimum user's energy consumption, the problem can be formulated  as below:
\begin{equation*}({\bf P2})\qquad
\begin{aligned}
\min_ {\ell }   \quad&  (L-\ell) C P_{\rm{cyc}}+ E_{\rm{off}}(\ell)
\qquad\rm{s.t.}~~ 
& \ell_{\min}^+ \le \ell \le L,
\end{aligned}
\end{equation*}
where $\ell_{\min}^+$ enforces the deadline for local computing (see Section~\ref{Mod:SU}).
\subsection{Energy-Efficient Adaptive Offloading}\label{Sec:EgyPolicyOneShot}
In this sub-section, we present a tractable approach for solving the complex Problem P1, by defining an \emph{offloading feasibility tunnel}  and using it as the tool to derive the \emph{string-pulling} policy for the energy-efficient offloading.

First, one can observe that Problem P1 is feasible if and only if the offloaded data size is no larger than the maximum helper-CPU   resource (in bits), i.e., $\ell \le U_{{{\rm{bit}}}, K}$. To simplify the procedure, we first solve Problem P1 conditioned on the full-utilization of helper-CPU, namely $\ell=U_{{{\rm{bit}}}, K}$. Then, the solution is modified for the case of underutilization, namely $\ell<U_{{{\rm{bit}}}, K}$.

\subsubsection{Full-Utilization of Helper-CPU \emph{[$\ell = U_{{{\rm{bit}}}, K}$]}} \label{Sec:FullCPU}The design approach consists of  constructing an offloading feasibility tunnel and pulling a  string (shortest path) over the tunnel as follows.
\noindent\underline{\textbf{a) Offloading Feasibility Tunnel}}\\
 To define the tunnel, we first derive two sets of constraints that specify the ceiling and floor of the tunnel. For the current case, one  key observation is that to meet the deadline constraint, the feasible solution should utilize all the  helper-CPU idle epochs. Mathematically, this introduces a set of \emph{helper-CPU computing-speed constraints} on the computed bits in each epoch $d_k(\ell_k)$ as:
\begin{equation}\label{Eq:Simplification}
d_k(\ell_k)\!=\!\dfrac{a_k \tau_k f_h}{C} \le B_{k-1}+\ell_k,~~~\text{and}~~~\sum_{j=1}^{k} d_j(\ell_j)=U_{{{\rm{bit}}},k}, \qquad k=1,\cdots, K.
\end{equation}
Combining \eqref{Eq:Simplification} with the remaining bits for computing, namely $B_k\!=\!\sum_{j=1}^k \ell_j\!-\!\sum_{j=1}^{k} d_j(\ell_j)$, yields 
\begin{equation}\label{Eq:LowerBoundary}
\text{(Minimum accumulated offloaded data size)}~~~
 \sum_{j=1}^k \ell_j \ge U_{{{\rm{bit}}},k} , \qquad k=1,\cdots, K.  
\end{equation}
Each of the above  constraints specifies the \emph{minimum accumulated offloaded data size} at a particular time instant. Next, substituting the helper-CPU computing-speed constraints in \eqref{Eq:Simplification} into the buffer constraint in \eqref{Eq:Buffer} leads to
 \begin{equation}\label{Eq:UpperBoundary}
 \text{(Maximum accumulated offloaded data size)}~~~
\sum_{j=1}^k \ell_j \le \min\{U_{\rm{bit},k}+Q, \ell\}, ~~ k=1,\cdots, K, 
 \end{equation} 
 which imposes the \emph{maximum accumulated offloaded data size} at each time instant.

Let an offloading policy be specified by a sequence of offloaded bits for different epochs: $\boldsymbol{\ell}=[\ell_1, \ell_2, \cdots, \ell_{K}]$. Then the \emph{offloading feasibility tunnel} is defined as follows.
\begin{definition}[Offloading Feasibility Tunnel]\emph{
Let $\mathcal{T}(\ell)$ denote the offloading feasibility tunnel for the total offloaded data size $\ell$, defined as the set of feasible offloading policies under the constraints in \eqref{Eq:LowerBoundary}, \eqref{Eq:UpperBoundary} and the deadline. Mathematically,
%\newpage
\begin{align}\label{Eq:OffloadingFeaTunnel}
%&\text{(Offloading Feasibility Region)}\nn\\
&\mathcal{T}(\ell)\!=\!\! \l\{ \boldsymbol{\ell} ~\bigg|~U_{{\rm{bit}},k} \!\le\! \sum_{j=1}^k \ell_j\le \min\{U_{{\rm{bit}},k}+Q, \ell\}, 
 ~\text{for}~ k\!=\!1,\cdots, K-1, ~\text{and}~ \sum_{k=1}^{K} \ell_k \!=\! \ell\r\}.
\end{align}}
\end{definition}

Graphically, the set of constraints in \eqref{Eq:LowerBoundary} depicts the floor of the tunnel and that in \eqref{Eq:UpperBoundary} its celling. Since constant-rate transmission within each epoch is optimal, the definition of the offloading feasibility tunnel can be plotted in the plane of number of bits versus time as illustrated in Fig.~\ref{Fig:OptimalScheduling}. One can observe that the tunnel floor is the helper's CPU-idling profile and shifting the floor upwards by the buffer size gives the tunnel ceiling. Specifically, for the case where the helper has a \emph{large buffer} for storing the offloaded data, referring to the case where $Q\ge L$, we have the following remark.
%the corresponding offloading feasibility tunnel can be further reduced as shown in the remark below.
\begin{remark}[Offloading Feasibility Tunnel for Large Buffer]\label{Rem:OffTunLarge}\emph{Consider that the helper has a large buffer. It has $\sum_{j=1}^k \ell_j\le \min\{U_{{\rm{bit}},k}+Q, \ell\}=\ell$, and thus the corresponding offloading feasibility tunnel can be reduced to the one that has a ceiling bounded by the total offloaded data size $\ell$ and the same floor as that of \eqref{Eq:OffloadingFeaTunnel}. Mathematically,
\begin{align}\label{Eq:OffloadingFeaTunnelLarge}
%&\text{(Offloading Feasibility Region)}\nn\\
&\mathcal{T}(\ell)\!=\!\! \l\{ \boldsymbol{\ell} ~\bigg|~U_{{\rm{bit}},k} \!\le\! \sum_{j=1}^k \ell_j\le  \ell, 
 ~\text{for}~ k\!=\!1,\cdots, K-1, ~\text{and}~ \sum_{k=1}^{K} \ell_k \!=\! \ell\r\}.
\end{align}}
\end{remark}
 \begin{figure}[t]
\begin{center}
\includegraphics[height=5.6cm]{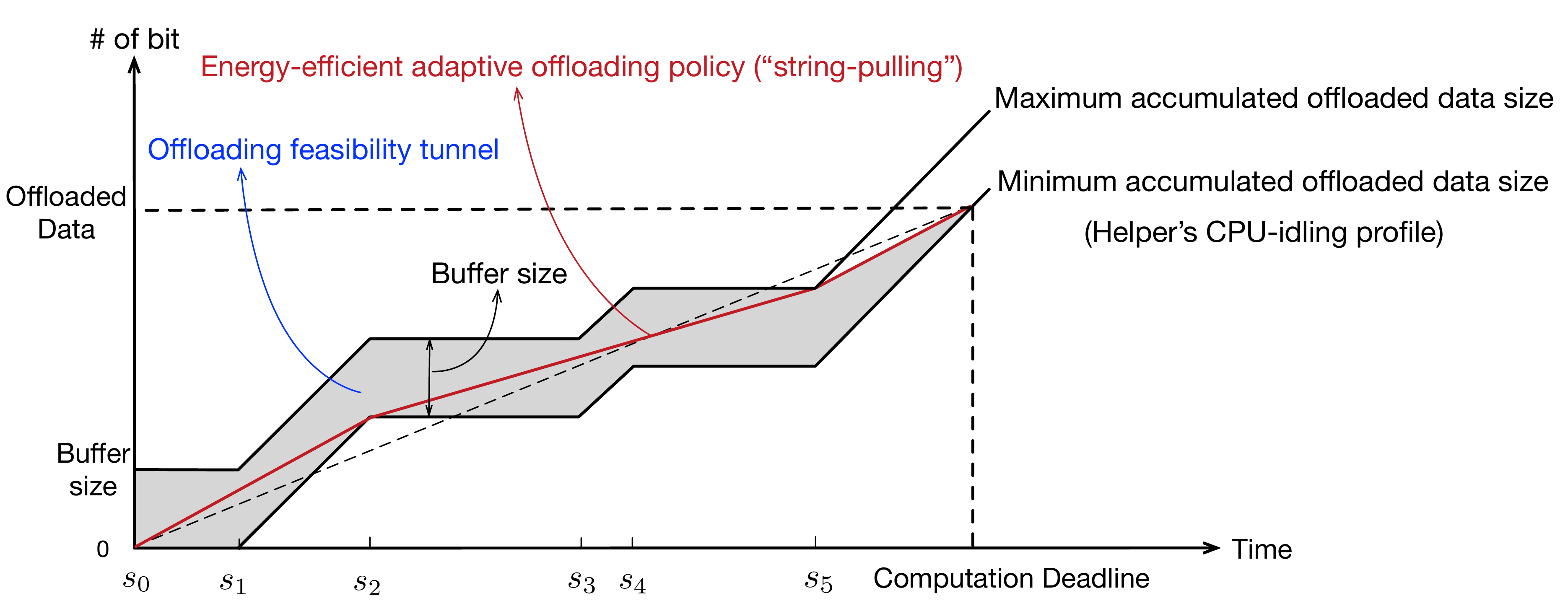}
\caption{An offloading feasibility tunnel (shaded in gray) and the energy-efficient transmission policy (the ``pulled string" in red) for the case of a small buffer at the helper.}
\label{Fig:OptimalScheduling}
\end{center}
\end{figure}

Using the said tunnel, Problem P1 can be equivalently transformed into Problem P3 below.
\begin{equation*}
({\bf P3})\qquad
\begin{aligned}
\min_ {\boldsymbol{\ell}\ge \boldsymbol{0}}   \quad& \sum_{k=1}^{K} \dfrac{\tau_k}{h^2} f\!\l(\dfrac{\ell_k}{\tau_k}\r)  
\qquad \rm{s.t.}~~ 
& \boldsymbol{\ell} \in \mathcal{T}(\ell).
\end{aligned}
\end{equation*}
It is easy to prove that Problem P3 is a convex optimization problem which can be solved by the Lagrange method. Instead, using the defined offloading feasibility tunnel, we show that the optimal policy has a ``string-pulling" structure in the sequel. Before the derivation, we define the ``string-pulling" policy and offer a remark introducing its application in transmission control. 
\begin{definition}[``String-Pulling" Policy]\label{Def:StrPul}\emph{Given a tunnel with a floor and ceiling (see Fig.~\ref{Fig:OptimalScheduling}), the ``string-pulling" policy is a scheme to construct the \emph{shortest path} from a starting point to an ending point through the tunnel, which can be performed by pulling  a stretched string from the same starting point to the ending point through the tunnel.}
\end{definition}
\begin{remark}[General ``String-Pulling" Transmission Control]\label{Rem:SP}\emph{The well-know class of ``string-pulling" policies for adapting transmissions arises from two simple facts. 
\begin{enumerate}
\item The transmission energy is a \emph{convex  increasing} function of the rate. 
\item Given data and time duration, the \emph{constant-rate transmission} is energy-efficient, corresponding to a straight-line segment in the throughput-time plane.
\end{enumerate} 
Time-varying transmission constraints, such as random energy or data arrivals, create a feasibility tunnel in the said plane. Given the above facts, a control policy for energy-efficient or throughput-optimal transmission is reflected as a ``pulled string" across the tunnel\cite{yang2012optimal,tutuncuoglu2012optimum,zafer2005calculus}.}
\end{remark}
\noindent\underline{\textbf{b) Optimal String-Pulling Policy}}\\
The offloading policy that specifies the set of offloaded bits for solving Problem P3 is shown to belong to the class of ``string-pulling" policies  in the  following proposition.

\begin{proposition}[Energy-Efficient Offloading Policy]\label{Prop:EgMinCondi}\emph{In the offloading feasibility tunnel $\mathcal{T}(\ell)$, the energy-efficient transmission policy $\boldsymbol{\ell}^*$ can be derived by forming a \emph{shortest path} connecting the starting and ending points, i.e., $(0, 0)$ and $(T_{\rm{end}}, \ell)$. Specifically, $\ell_k^*\!=\! r_k^* \tau_k$ where  the optimal offloading rate in each epoch, $r_k^*$, satisfies the following necessary and sufficient conditions.
\begin{itemize}
\item[1)]{The offloading  rate  does not change unless the helper-CPU  changes its state.}
\item[2)]{The offloading rate can increase only when the helper-CPU  changes to enter the idle state. In this case, the buffer is fully occupied at the time instant, e.g., at time $s_k$, it has $B_{k-1}=Q$.}
\item[3)]{The offloading rate can decrease only when the helper-CPU  changes to enter  the busy state. In this case, the buffer is empty at the time instant, e.g., at time $s_{k}$, it has $B_{k-1}=0$.}
\end{itemize}
}
\end{proposition}

The shortest path in Proposition~\ref{Prop:EgMinCondi} can be derived by the ``string-pulling" policy defined in Definition~\ref{Def:StrPul} and illustrated in Fig.~\ref{Fig:OptimalScheduling}. The result can be proved by showing  that the optimal offloading policy satisfies the facts specified in Remark~\ref{Rem:SP}. The details are straightforward and omitted for brevity.

For computing the policy, There exist a generic algorithm for finding a shortest path across a tunnel as in Proposition~\ref{Prop:EgMinCondi} \cite{tutuncuoglu2012optimum, zafer2005calculus}. The key idea is to \emph{recursively} find the set of tuning time instants and the slopes of path segments (offloading rates in the current case) between the time instants. The algorithm can be modified for computing the energy-efficient offloading policy. To this end, several definitions are needed. Consider the starting time-instant $s_i$ where $i$ is the epoch index. Define the \emph{reference rate region} of epoch $k$ by $\boldsymbol{R}_k=\{r_k | R_{k}^{\min}\le r_k\le R_{k}^{\max}\}$ for $s_i+1\le k\le K $, where  
\begin{align}
R_{k}^{\min}=\dfrac{U_{{\rm{bit}},k}-\sum_{j=1}^i \ell_j^*}{s_k-s_i}~~\text{and}~~
  R_{k}^{\max}=\dfrac{\min\{U_{{\rm{bit}},k}+Q, \ell\}-\sum_{j=1}^i \ell_j^*}{s_k-s_i}
\end{align} are the minimum and maximum offloading rate that result in the empty and fully-occupied buffer at the end of epoch $k$, respectively. Note that when $i=0$, $\sum_{j=1}^i \ell_j^*=0$ and when $i>0$, $\ell_j^*$ for $1\le j\le i$ is the decided offloaded data size.  In addition, define $R_i^{\rm{end}}$ as the \emph{reference constant rate} at time instant $s_i$, given as $R_i^{\rm{end}}=\frac{\ell-\sum_{j=1}^i \ell_j^*}{T_{\rm{end}}-s_i}$, corresponding to the slope of a straight-line connecting the staring and ending points.
Note that the rates \{$R_{k}^{\min}$, $R_{k}^{\max}$, $R_i^{\rm{end}}$\} may not be feasible but used for comparison. The detailed algorithm is presented in Algorithm~\ref{Alg:P4}.
\begin{algorithm}[t!]
  \caption{Computing the Energy-Efficient Offloading Policy for Solving Problem P3.}
  \label{Alg:P4}
  \begin{itemize}
\item{\textbf{Step 1} [Initialization]:  $n=1$, $i_n^s=1$ and $k=i_n^s$ where $i_n^s$ is the epoch index of the starting time instant for the $n$-th constant rate.
 }
\item{\textbf{Step 2} [Determine the ``string-pulling" offloading policy]:\\  
(1) Check whether the reference rate $R_{i_n^s}^{\rm{end}}$ is feasible: if $R_{i_n^s}^{\rm{end}}\in \bigcap_{j=i_n^s}^{K} \boldsymbol{R}_j$, transmit at rate $R_{i_n^s}^{\rm{end}}$ from epoch $i_n^s$ to $K$ and terminate the algorithm, otherwise, go to the next sub-step.
 \\(2) Find the next turning time instant of the shortest path and compute the offloading rate:
 \begin{itemize}
 \item[i)]{While $\boldsymbol{R}_{k+1}\in \bigcap_{j=i_n^s}^k \boldsymbol{R}_j$, update by $k=k+1$, otherwise, go to the next sub-step.}
 \item[ii)]{If $\boldsymbol{R}_{k+1}> \bigcap_{j=i_n^s}^k \boldsymbol{R}_j$, then $i_n^e=m$ where $m=\max\{j~|~R_{j}^{\max}=r_k^* \}$ and  $r_k^*=\max\{\bigcap_{j=1}^k \boldsymbol{R}_j\}$. For $i_n^s \le k\le i_n^e$, the optimal offloaded data size is $\ell_k^*=r_k^* \tau_k$ .  \\
 If $\boldsymbol{R}_{k+1}< \bigcap_{j=i_n^s}^k \boldsymbol{R}_j$, then $i_n^e=m$ where $m=\max\{j~|~R_{j}^{\min}=r_k^* \}$ and  $r_k^*=\min\{\bigcap_{j=1}^k \boldsymbol{R}_j\}$. For $i_n^s \le k\le i_n^e$, the optimal offloaded data size is $\ell_k^*=r_k^* \tau_k$ .}
 \end{itemize}
\item{\textbf{Step 3} [Repeat]: Let $n=n+1$, $i_n^s=i_{n-1}^e+1$, $k=i_n^s$;
update $\boldsymbol{R}_k$ and go to Step 2. }
}
\end{itemize}
  \end{algorithm}
  
   \begin{remark}[Buffer Gain]\label{Rem:BufferGain}\emph{ It can be observed from  Fig.~\ref{Fig:OptimalScheduling} that increasing the buffer size will shift up the tunnel ceiling, enlarging the tunnel area. This allows the pulled string to approach a single straight-line and thereby reduce the transmission-energy consumption. However, the buffer gain saturates when the buffer size exceeds the total offloaded bits, corresponding to a large buffer.
}
\end{remark}

 \begin{remark}[Effect of Helper's CPU-Idling Profile]\emph{It can be observed from Fig.~\ref{Fig:OptimalScheduling} that the helper's CPU-idling profile significantly affects the energy-efficient P2P transmission policy. Specifically, when the helper has a large buffer, the optimal offloading policy is only constrained by the tunnel floor (see \eqref{Eq:OffloadingFeaTunnelLarge}). Given total helper-CPU idling duration, the user can achieve the minimum transmission-energy consumption if the helper's CPU  first stays at the busy state and then switches to the idle state that lasts until the deadline. The reason is that in this scenario, the user has a long consecutive duration for transmitting enough input data for fully utilizing helper-CPU idle epochs, resulting in low transmission rates.}
\end{remark}

\subsubsection{Underutilization of Helper-CPU \emph{[$\ell < U_{{{\rm{bit}}}, K}$]}}  \label{Sec:UnderutilizationEgy} This case is desirable  in two scenarios. First, the spare CPU resource at the helper is rich such that its full utilization  may not be necessary or even possible. Second, when the channel is unfavorable, it is beneficial to reduce the offloaded data size which may under-utilize the helper's CPU. To characterize the corresponding policy structures, in the following, we first consider the large buffer case and derive its optimal offloading policy. While for the case of small buffer, the corresponding problem is highly complex. To address this challenge, we design a sub-optimal  policy using the insight from the  large buffer counterpart. \\
\noindent\underline{\textbf{a) Large Buffer}}\\
Consider that the helper has a large buffer (i.e.,  $Q\ge L$). For the case of underutilization of helper-CPU, the offloaded bits $\ell$ is below the  helper's spare CPU capacity. The corresponding optimal offloading strategy can be designed by extending the solution approach for the full-utilization counterpart. This essentially involves defining an \emph{effective offloading feasibility tunnel} with a lower floor with respect to (w.r.t.) the original one in \eqref{Eq:OffloadingFeaTunnelLarge}. See the details below.

 Recall the CPU real-time constraints, namely that a CPU cycle available now cannot be used in the future. Then given the helper's CPU-idling profile  $U_{\text{bit}, K}$ and offloaded data bits for computing $\ell$, the amount of underutilized CPU resource, measured by the accumulated unused computable bits in each epoch, cannot exceed $\Delta(\ell)=(U_{{{\rm{bit}}}, K}-\ell)$-bit. Otherwise, computing the $\ell$-bit of offloaded data by the deadline is infeasible. Mathematically, 
\begin{equation*}
U_{{{\rm{bit}}},k}-\sum_{j=1}^{k} d_j(\ell_j)\le \Delta(\ell),~\text{for}~ k=1, \cdots, K-1,~~\text{and}~~ U_{{{\rm{bit}}},K}-\sum_{j=1}^{K} d_j(\ell_j)=\Delta(\ell)
\end{equation*}  where $d_j(\ell_j)$ gives the bits computed in epoch $j$ as defined earlier. Combing the constraints with the property of accumulated computed bits: $0\le\! \sum_{j=1}^{k} d_j(\ell_j)\le \min\l\{U_{{{{\rm{bit}}}},k},  \sum_{j=1}^{k} \ell_j\r\},$ which can be observed from \eqref{Eq:BEvolve}, yields the bounds on the accumulated computed bits below: 
\begin{equation}\label{Eq:NewComputed}
[U_{{{\rm{bit}}},k}-\Delta(\ell)]^{+} \le \sum_{j=1}^k d_j(\ell_j)\le \min\l\{U_{{{{\rm{bit}}}},k},  \sum_{j=1}^{k} \ell_j\r\},  ~k=1,\cdots, K.
\end{equation}
Using  \eqref{Eq:NewComputed}, the \emph{effective offloading feasibility tunnel} is defined as follows.
\begin{definition}[Effective Offloading Feasibility Tunnel]\emph{Assume that the helper has a large buffer. For the case of underutilization, the effective offloading feasibility tunnel, denote by $\bar{\mathcal{T}}(\ell)$,  is defined as the set of policies with accumulated offloaded bits constrained as 
\begin{align}\label{Eq:EffectiveRegion}
&\bar{\mathcal{T}}(\ell)=\l\{ \boldsymbol{\ell}  ~\bigg|~[U_{{{\rm{bit}}},k}-\Delta(\ell)]^{+} \le \sum_{j=1}^{k} \ell_j\le \ell , ~ \text{for}~k=1,\cdots, K-1,~\text{and}~ \sum_{k=1}^{K} \ell_k = \ell~\r\}.
\end{align}}
\end{definition}
The effective offloading feasibility tunnel can be constructed by shifting downwards the full-utilization counterpart  $\mathcal{T}(\ell)$ in \eqref{Eq:OffloadingFeaTunnelLarge} by $(U_{{{\rm{bit}}}, K}-\ell)$ and then cutting regions where the number of bits is below $0$. 
Next, one important property of the defined effective offloading feasibility tunnel is stated in the proposition below, proved in Appendix~\ref{App:OptRegion}.
% the extended version \cite{you2017energy}. 
%Appendix~\ref{App:OptRegion}.
\begin{proposition}\label{Pro:OptRegion}\emph{Assume that the helper has a large buffer. For the case of underutilization, the energy-efficient transmission policy can be derived by forming a shortest path in the effective offloading feasibility tunnel.
}
\end{proposition}
Based on Proposition~\ref{Pro:OptRegion}, Problem P1 can be transformed into the problem with constraints replaced by the effective offloading feasibility tunnel. The new problem has the same form as Problem P3 and only differs in the definitions of offloading feasibility tunnel. Thus, it can be solved using the same ``string-pulling" approach as in Section~\ref{Sec:FullCPU}.

\noindent\underline{\textbf{b) Small Buffer}}\\
For this case, we show that computing the optimal policy is highly complex without yielding useful insight. To address this difficulty, we propose a tractable \emph{proportional CPU-utilization} scheme which is asymptotically optimal.

First, similar to the case of large buffer, 
%Section~\ref{Sec:UnderutilizationEgy}, 
given the helper's CPU-idling profile and the deadline constraint, the amount of unused computable bits is $\Delta(\ell)=(U_{{{\rm{bit}}}, K}-\ell)$-bit and the accumulated computed bis can be bounded as \eqref{Eq:NewComputed}. Combining them with the buffer constraints in \eqref{Eq:Buffer} yields the following constraints on the accumulated offloaded bits:
\begin{equation}
 \sum_{j=1}^k d_j(\ell_j) \le \sum_{j=1}^{k} \ell_j\le \sum_{j=1}^k d_j(\ell_j)+Q,~~k=1, \cdots K.
\end{equation} Therefore, Problem P3 can be transformed into Problem P4 as follows.
\begin{equation*}~~({\bf P4})~
\begin{aligned}
\min_ {\boldsymbol{\ell}\ge \boldsymbol{0}}   \quad& \sum_{k=1}^{K} \dfrac{\tau_k}{h^2} f\!\l(\dfrac{\ell_k}{\tau_k}\r) \\
\rm{s.t.}\quad & \sum_{j=1}^k d_j(\ell_j) \le \sum_{j=1}^{k} \ell_j\le \sum_{j=1}^k d_j(\ell_j)+Q, \quad k=1,\cdots, K,
\\ & [U_{{{\rm{bit}}},k}-\Delta(\ell)]^{+} \le \sum_{j=1}^k d_j(\ell_j)\le U_{{{{\rm{bit}}}},k}, \quad k=1,\cdots, K,\\
& \sum_{k=1}^{K} d_k(\ell_k)=\sum_{k=1}^{K} \ell_k=\ell.
\end{aligned}
\end{equation*}
Since $d_j(\ell_j)$ is a non-affine function of $\ell_j$ (see \eqref{Eq:BEvolve}), Problem P4 is a non-convex optimization problem that is difficult to solve. The intractability arises from determining the time instants and levels (in terms of unused CPU cycles) the helper-CPU should be under-utilized, which are coupled due to residual unused CPU resource delivered from one epoch to the next. The conventional approach for solving this type of optimization problem is using dynamic programming, which requires discretizing the continuous state space, bringing high complexity but without yielding useful insight on the policy structures. To tackle the difficulty, we propose the following practical scheme of \emph{proportional CPU-utilization}.

%For this case (i.e., $Q<\ell$), computing the optimal policy requires solving a non-convex problem (see the extended version \cite{you2017energy}). The intractability arises from determining the time instants and levels (in terms of unused CPU cycles) the helper-CPU should be under-utilized, and the policy should satisfy the CPU real-time constraints. To overcome the difficulty, we propose the following practical scheme of \emph{proportional CPU-utilization}.

\begin{definition}[Proportional CPU-Utilization]\label{Def:ProCPUUtili}\emph{Consider the helper has a small buffer. For the case of underutilization, in each CPU idle epoch, the proportional CPU-utilization scheme assigns a fixed number of CPU cycles to the user per second without adjusting the CPU frequency. As a result, the user can fully utilize the allocated CPU resource. Let $\tilde{f}_h$ denote the number of allocated CPU cycles per second. Mathematically, $\tilde{f}_h=f_h \frac{\ell}{U_{{\rm{bit}},K}}$. }
\end{definition}
This scheme can be implemented by the advanced hyper-threading technique \cite{koufaty2003hyperthreading} which allows multi-thread to time-share one physical CPU via proportional CPU resource allocation. Under this scheme, we define $\tilde{U}_{{\rm{bit}},k}$ as an \emph{effective helper's CPU-idling profile}, give as $\tilde{U}_{{\rm{bit}},k}=U_{{\rm{bit}},k}\frac{\ell}{U_{{\rm{bit}},K}}$, for $k=1, \cdots, K$.
Then the current case of underutilization of helper-CPU can be reduced to the counterpart of full-utilization in Section~\ref{Sec:FullCPU} and  efficiently solved using the same approach.  Furthermore, this scheme is shown to be \emph{asymptotically optimal} in the following proposition.
\begin{proposition}[Asymptotic Optimality]\label{Prop:ProCPUUtil}\emph{The proportional CPU-utilization scheme is the optimal offloading policy when the buffer size $Q\to 0$.}
\end{proposition}
This proposition is proved in Appendix~\ref{App:ProCPUUtil}. It indicates that when the buffer size is smaller, the performance of proposed scheme gets closer to that of the optimal one.
}}

\subsection{Energy-Efficient Data Partitioning}\label{Sec:OptDataEgy}
The direct derivation for energy-efficient data partitioning in Problem P2 is intractable due to the lack of closed-form expression for the minimum transmission-energy consumption, i.e., $E_{\rm{off}}(\ell)$, which can be observed from Proposition~\ref{Prop:EgMinCondi}. To overcome this difficulty, in this sub-section, Problem P2 is proved to be a convex optimization problem, allowing the optimal solution to be computed by a sub-gradient method.

First, to guarantee that both the adaptive offloading and local computing are feasible, the offloaded data bits should satisfy: $\ell_{\min}^{+}\le \ell \le \min\l\{U_{{\rm{bit}}, K}, L\r\}.$ Therefore, Problem P2 is feasible if and only if $\ell_{\min}^+\le U_{{\rm{bit}}, K}$.  Next, let $\ell^{(1)}$ and $\ell^{(2)}$ denote two offloaded data bits. Since the offloading feasibility tunnel $\mathcal{T}(\ell)$ in \eqref{Eq:OffloadingFeaTunnel} can be regarded as one special case of $\bar{\mathcal{T}}(\ell)$ in \eqref{Eq:EffectiveRegion} for which $\ell=U_{{\rm{bit}},K}$, we only consider the effective offloading feasibility tunnel in this sub-section. One important property of  the tunnel is presented below, proved in 
%the extended version \cite{you2017energy}. 
Appendix~\ref{App:PropRegi}.
\begin{lemma}\label{Lem:PropRegi}\emph{
Let $\boldsymbol{\ell}^{(1)} \in \bar{\mathcal{T}}(\ell^{(1)})$ and $\boldsymbol{\ell}^{(2)} \in \bar{\mathcal{T}}(\ell^{(2)})$. Then, for $0\le \lambda \le 1$,
\begin{equation}
\lambda \boldsymbol{\ell}^{(1)} + (1-\lambda) \boldsymbol{\ell}^{(2)}  \in \bar{\mathcal{T}} (\lambda \ell^{(1)} + (1-\lambda) \ell^{(2)}).
\end{equation}
}
\end{lemma}

Using Lemma~\ref{Lem:PropRegi}, the convexity of the function $E_{\rm{off}}(\ell)$ is stated in the following lemma.
 
\begin{lemma}[Convexity of Minimum Transmission-Energy Function]\label{Lem:MasConvTime}\emph{
The function of minimum transmission-energy consumption, $E_{\rm{off}}(\ell)$, is a convex function w.r.t $\ell$. }
\end{lemma}

Lemma~\ref{Lem:MasConvTime} is proved in Appendix~\ref{App:MasConvTime}.
Using this lemma, it can be easily verified that Problem P2 is a convex optimization problem. Directly applying KKT conditions yields the key result of this sub-section in the proposition below.
\begin{proposition}[Energy-Efficient Data Partitioning]\label{Prop:OptDivisionEgy}\emph{Given the computation load $L$ and deadline $T$ at the user, the energy-efficient data-partitioning policy solving Problem P2 is:
\begin{equation*}
\ell^*=\max\l\{\ell_{\min}^+, \min\l\{\ell_0,U_{{\rm{bit}},K}, L\r\}\r\}
\end{equation*}
where $\ell_0$ is the solution for $E_{\rm{off}}^{'}(\ell_0)=C P_{\rm{cyc}}$ and $E_{\rm{off}}^{'}(\ell)$ denotes the first derivative of $E_{\rm{off}}(\ell)$.
}
\end{proposition}
Although the function $E_{\rm{off}}^{'}(\ell)$ has no closed form, $\ell_0$ in Proposition~\ref{Prop:OptDivisionEgy} can be easily computed via advanced convex optimization techniques, e.g., the sub-gradient method, yielding the optimal data partitioning using the formula in the proposition.

{\color{black}{Last,  $E_{\rm{off}}(\ell)$ can be lower-bounded as $E_{\rm{off}}(\ell)\ge \frac{T_{\rm{end}}}{h^2} f(\ell/T_{\rm{end}})$. Combining it with Proposition~\ref{Prop:OptDivisionEgy} gives the following corollary.
\begin{corollary}[Minimum Offloading]\emph{Given the computation load $L$ and deadline $T$ at the user, if it satisfies that $T_{\rm{end}} f^{-1}(C_mP_{\rm{cyc}}h^2)\le \ell_{\min}^+$, the energy-efficient data partitioning selects the minimum data size for offloading, i.e., $\ell^*=\ell_{\min}^+$.}
\end{corollary}
This corollary reduces the complexity for computing the data partitioning policy if the said condition is satisfied. Moreover, it is coincident with the intuition that if the user has a bad channel or local computing consumes small energy, it is preferred to reduce the offloaded bits.
}}
{\color{black}{
\begin{remark}[Offloading to Multiple Helpers]\emph{The current results can be extended to the case where the user can offload input data to multiple helpers. The corresponding design can be formulated as a hierarchical optimization problem.  Specifically, the slave problem aims at minimizing the energy consumption for offloading to one particular helper, for which the optimal offloading policy can be derived by the same ``string-pulling" approach. On the other hand, the master problem focuses on partitioning input data for local computing and offloading to multiple helpers. This optimization problem can be proved to be also convex  using Lemma~\ref{Lem:MasConvTime}, thus the optimal data partitioning policy can be computed by the sub-gradient method.}
\end{remark}}
\begin{remark}[Co-Computing Networks]\emph{Our current design can be used as a building block for implementing different types of networks such as multi-helper networks and multi-access networks. For multi-helper networks, the helper selection can be performed as follows. Assume each user selects one helper that is within a certain distance and has the largest amount of idling computation resource given the deadline. Once the cooperation is initiated, the helper is assumed to be dedicated for co-computing with this user until the deadline. Next, consider multi-access networks where multiple users offload computation to one helper. The designs of adaptive offloading and data partitioning can be integrated with computation resource allocation at the helper such as the proposed proportional CPU-utilization scheme (see Definition~\ref{Def:ProCPUUtili}).}
\end{remark}}

\section{Mobile Cooperative Computing with Bursty Data Arrivals}\label{Sec:RandomComputLoad}
In this section, the solution approach for energy-efficient co-computing as developed in Section~\ref{Sec:EgyOneShot} is extended to the case of bursty data arrivals. The data bursty introduces a set of so-called \emph{data causality constraint} defined in the sequel. Due to the new constraints, the original algorithms for offloading and data partitioning need be redesigned. This essentially involves defining an alternative offloading feasibility tunnel accounting for bursty data arrivals. 

\subsection{Problem Formulation}
Consider the user has bursty data arrivals at time instants $\!\{\hat{s}_k\}$ as shown in Fig.~\ref{Fig:SecPri} and the helper has a large buffer (i.e., $Q \!\ge\! \!\sum_{k=1}^K L_k$)\footnote{Note that $T_{\rm{end}}=T$ if the last epoch of the helper-CPU idling profile is idle. Moreover, the extension to the case of small buffer can be modified from those for the large buffer case using the similar approach for the one-shot arrival counterpart, and thus omitted for brevity.}. 
{\color{black}{Allowing each instant of data arrivals to have different partitioning ratios makes the optimization problem intractable without yielding useful insight.  To tackle this challenge, we first propose a tractable \emph{proportional data partitioning} scheme as defined below, which allows using the similar ``string-pulling" approach in the sequel.}}
\begin{definition}[Proportional Data Partitioning]\emph{For the $k$-th event time-instant $\hat{s}_k$, let $L_{k,\rm{off}}$ denote the size of partitioned data for offloading. The scheme of proportional data partitioning divides the data of each arrival for local computing and offloading using a fixed ratio: $\dfrac{L_{1,\rm{off}}}{L_1}=\dfrac{L_{2,\rm{off}}}{L_2}=\cdots=\dfrac{L_{K,\rm{off}}}{L_K}=\theta$, where $\theta$ is called the \emph{data-partitioning ratio}.}
\end{definition}
Note that when there is no data arrival at time instant $\hat{s}_k$, $L_k=0$ (see Section~\ref{Mod:SU}). The data-partitioning ratio $\theta$ is the optimization variable in the problem of data partitioning.

Based on the above definition, the problem of energy-efficient co-computing for bursty data arrivals can be decomposed as the following slave and master problems.

\subsubsection{Slave Problem of Adaptive Offloading} First, we derive a set of data causality constraints arising from bursty data arrivals. They reflect the simple fact: an input-data bit cannot be offloaded or computed before it arrives. Equivalently, for each event time-instant $\hat{s}_k$, the user partitions $(\theta L_k)$-bit data for offloading given a fixed data-partitioning ratio $\theta$. The accumulated offloaded data size cannot exceed size of the $\theta$-fraction of the accumulated data size for every time instant. Mathematically, 
\begin{equation}\label{Eq:DataCausalityOff}
\text{(Data causality constraints for offloading)} \quad \sum_{j=1}^k \ell_j \le \sum_{j=1}^{k-1} \theta L_j, \quad k=1,\cdots, K.~~~  
\end{equation}
\begin{remark}[Similarities with Energy-Harvesting Transmissions]\emph{
The data causality constraints are analogous with the energy causality constraints for energy-harvesting transmissions \cite{tutuncuoglu2012optimum,OzelUlukus:TransEnergyHarvestFading:OptimalPolicies:2011}. The latter specify that the accumulated energy consumed by transmission cannot exceed the total harvested energy by any time instant. The data constraints are due to random data arrivals while the energy counterparts arise from random energy arrivals. The above analogy together with that in Remark~\ref{Rem:SP} establish an interesting connection between the problem mathematical structures in the two different areas: energy-harvesting communications and co-computing.}
\end{remark}
By modifying Problem P1 to include the above constraints and assuming large buffer, the problem of 
energy-efficient offloading is formulated as:
\begin{equation*}({\bf P5})\qquad
\begin{aligned}
\min_ {\boldsymbol{\ell}\ge \boldsymbol{0} }   \quad& \sum_{k=1}^{K} \dfrac{\tau_k}{h^2} f\!\l(\dfrac{\ell_k}{\tau_k}\r) \\
\rm{s.t.}\quad 
& \sum_{k=1}^{K} d_k(\ell_k)=\sum_{k=1}^{K} \ell_k=\sum_{k=1}^{K-1} \theta L_k, \\ 
& \sum_{j=1}^k \ell_j \le \sum_{j=1}^{k-1} \theta L_j, &\quad k=1,\cdots, K.
\end{aligned}
\end{equation*}
Let $\hat{E}_{\rm{off}}(\theta)=\sum_{k=1}^{K} \dfrac{\tau_k}{h^2} f\!\l(\ell_k^*/\tau_k\r)$ denote the minimum transmission-energy consumption where $\{\ell_k^*\}$ solve Problem P5.
\subsubsection{Master Problem of Proportional Data Partitioning}
Given $\hat{E}_{\rm{off}}(\theta)$, the master problem focuses on optimizing the data-partitioning ratio $\theta$ under the criterion of the minimum user's energy consumption. Let $\ell_{\rm{loc},k}$ denote the size of data for local computing at the user  in epoch $k$. A set of data causality constraints for local computing can be derived similarly as \eqref{Eq:DataCausalityOff}: 
\begin{equation}\label{Eq:LocalComputingDataCausality}
\text{(Data causality constraints for local computing)}~ \sum_{j=1}^k \ell_{\rm{loc},j}\le \sum_{j=1}^{k-1} (1-\theta)L_j, ~ k=1,\cdots, \tilde{K}.
\end{equation}
Note that for local computing, it has $\tilde{K}$ epochs determined by the deadline $T$. Assume that the user's CPU performs local computing whenever there exists computable data or otherwise stays idle. Let   $d_{\rm{loc},k}(\ell_{\rm{loc},k})$ denote the bits computed locally in epoch $k$ and $B_{\rm{loc},k}$ the bits of remaining data at the end of epoch $k$. Due to the  CPU real-time constraints mentioned earlier, $d_{\rm{loc},k}(\ell_{\rm{loc},k})$ and $B_{\rm{loc},k}$ evolve as:
\begin{equation*}\label{Eq:LocalComputedData}
d_{\rm{loc},k}(\ell_{\rm{loc},k})\!=\! \min\l\{ B_{\rm{loc},k-1}+\ell_{\rm{loc},k}, \frac{ \tau_k f}{C}\r\} ~\text{and}~B_{\rm{loc},k}\!=\!\sum_{j=1}^k \ell_{\rm{loc},j}-\sum_{j=1}^{k} d_{\rm{loc},j}(\ell_{\rm{loc},j}), ~~ k=1,\cdots, \tilde{K},
\end{equation*} with $B_{\rm{loc},0}=0$. Under the data causality constraints in \eqref{Eq:LocalComputingDataCausality}, the problem of proportional data partitioning can be formulated as follows.
\begin{equation*}({\bf P6})\qquad
\begin{aligned}
\min_ {\theta, \boldsymbol{\ell}_{\rm{loc}}\ge \boldsymbol{0} }   \quad& \l[\sum_{k=1}^{\tilde{K}-1} (1-\theta) L_k\r] C P_{\rm{cyc}}+\hat{E}_{\rm{off}}(\theta)\\
\rm{s.t.}\quad 
& \sum_{k=1}^{\tilde{K}} \ell_{\rm{loc},k}= \sum_{k=1}^{\tilde{K}} d_{\rm{loc},k}(\ell_{\rm{loc},k})= \sum_{k=1}^{\tilde{K}-1} (1-\theta)L_k,\\
& \sum_{j=1}^k \ell_{\rm{loc},j}\le \sum_{j=1}^{k-1} (1-\theta)L_j, & k=1,\cdots, \tilde{K}. 
\end{aligned}
\end{equation*}
\subsection{Energy-Efficient Adaptive Offloading}\label{Sec:BurstyP2P}
In this sub-section,  the energy-efficient offloading policy is derive by defining an alternative offloading feasibility tunnel accounting for the bursty data arrivals.

The problem feasibility conditions are decided by the offloading feasibility tunnel summarized shortly. One of necessary conditions is that the total offloaded data is no larger than the helper's CPU resource, i.e., $\sum_{k=1}^{K-1} \theta L_k\le U_{{\rm{bit}},K}$. In the following, we solve Problem P5 conditioned on the full-utilization and underutilization of helper-CPU, respectively.
\subsubsection{Full-Utilization of Helper-CPU}
 The solution approach requires the definition of an offloading feasibility tunnel determined by the data causality constraints.
 
\begin{figure}[t]
\begin{center}
\includegraphics[height=5.6cm]{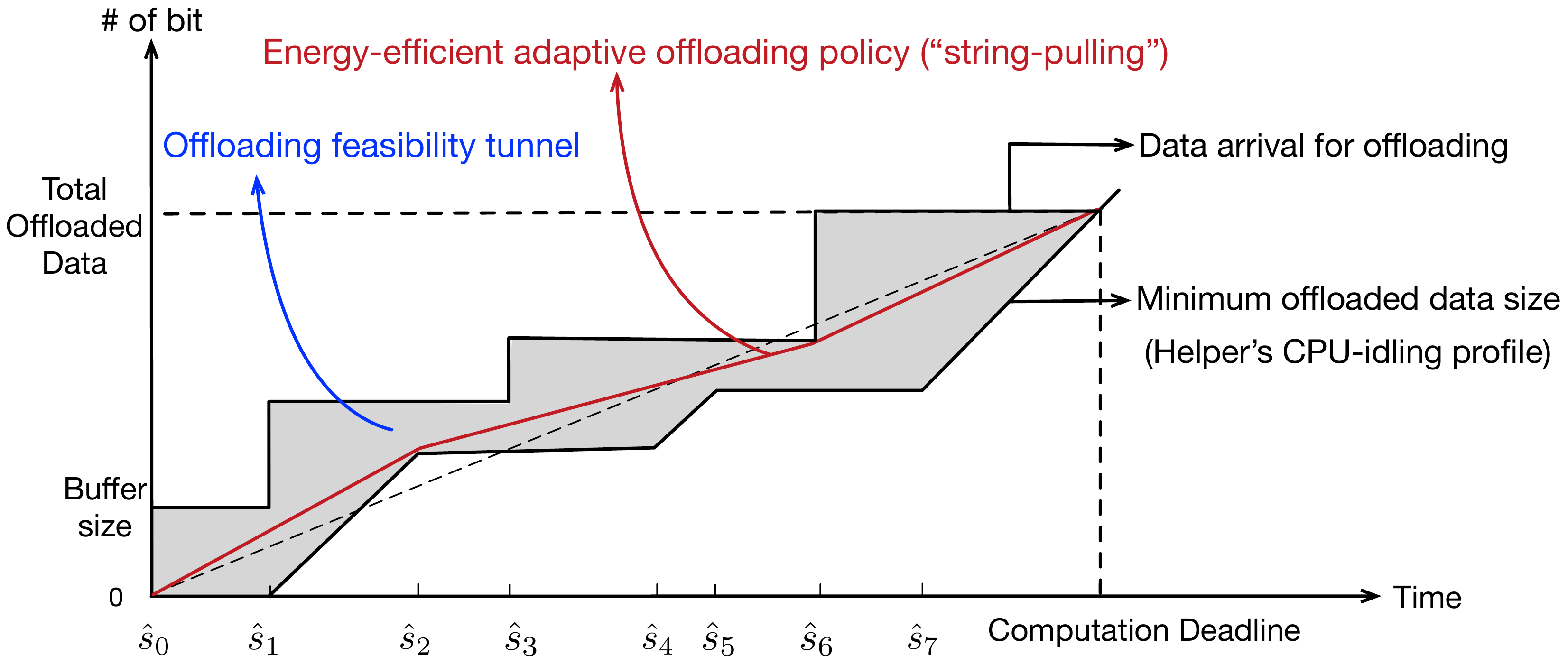}
\caption{An offloading feasibility tunnel (shaded in gray) for the case of bursty data arrivals and the energy-efficient transmission policy (the ``pulled string" in red).}
\label{Fig:BurstyOptimalScheduling}
\end{center}
\end{figure}

 To define the tunnel, we derive the conditions that specify the floor and ceiling of the tunnel. First, similar to Section~\ref{Sec:FullCPU}, the deadline constraint imposes the constraints on the minimum accumulated offloaded data size in \eqref{Eq:LowerBoundary}, specifying the tunnel floor. Next, the data causality constraints for offloading in \eqref{Eq:DataCausalityOff} determine the tunnel ceiling. Combing them together, we define the corresponding offloading feasibility tunnel as follows.
\begin{align}\label{Eq:OffFeaTunnel}
& \text{(Offloading Feasibility Tunnel for Bursty Data Arrivals)}\nn\\
&\mathcal{T}_B (\theta)\!=\!\l\{ \boldsymbol{\ell}  ~\bigg|~U_{{{\rm{bit}}},k} \le \sum_{j=1}^k \ell_j \le \sum_{j=1}^{k-1} \theta L_j,  ~\text{for}~ k=1,\cdots, K-1,~\text{and}~ \sum_{k=1}^{K} \ell_k \!=\!\sum_{k=1}^{K-1} \theta L_k\r\}.
\end{align}
The graphical illustration for the tunnel is given in Fig.~\ref{Fig:BurstyOptimalScheduling}. It suggests that Problem P5 is feasible if and only if the tunnel ceiling is always not below the tunnel floor. Mathematically, $U_{{{\rm{bit}}},k} \le \sum_{j=1}^{k-1} \theta L_j$, for $k=1, \cdots, K$. 
%Moreover, observe that compared with the offloading feasibility tunnel shown in Fig.~\ref{Fig:OptimalScheduling}, the current tunnel has a ceiling specifying the dynamics of bursty data arrivals for offloading. This can potentially increase the transmission-energy consumption (the length of the pulled string) due to extra constraints (the ceiling).

Given Problem P5 is feasible, it can be transformed to the one that replaces the constraints with the offloading feasibility tunnel. Again, the corresponding energy-efficient offloading policy can be computed by the said ``string-pulling" algorithm.
% \cite{you2017energy}.\\

 \subsubsection{Underutilization of Helper-CPU \emph{[$\ell < U_{{\rm{bit}}, K}$]}}  \label{Sec:UnderutilizationEgyBursty}
For this case, similar to the one-shot data arrival counterpart, the key step is to define an \emph{effective offloading feasibility tunnel}. 

Similar to Section~\ref{Sec:UnderutilizationEgy}, given the helper's CPU-idling profile and the deadline constraint, the amount of unused computable bits is $\bar{\Delta}(\theta) = (U_{{\rm{bit}},K}-\sum_{k=1}^{K-1} \theta L_k)$-bit and the accumulated computed bis can be bounded similar to that in \eqref{Eq:NewComputed}. Using \eqref{Eq:NewComputed} and the data causality constraints for offloading in \eqref{Eq:DataCausalityOff}, an effective offloading feasibility tunnel is defined as follows.
\begin{align}\label{Eq:EffectiveOffTunnel}
& \text{(Effective Offloading Feasibility Tunnel for Bursty Data Arrivals)}\nn\\
&\bar{\mathcal{T}}_B(\theta)=\l\{ \boldsymbol{\ell}  ~\bigg|~[U_{{{\rm{bit}}},k}-\bar{\Delta}(\theta)]^{+} \le \sum_{j=1}^k \ell_j \le \sum_{j=1}^{k-1} \theta L_j,  \right.\nn\\ &   ~~~~~~~~~~~~~~~~~~~~~~~~~~~~~~~~~\left.~\text{for}~ k=1,\cdots, K-1,~\text{and}~ \sum_{k=1}^{K} \ell_k =\sum_{k=1}^{K-1} \theta L_k\r\}.
\end{align}
Note that compared with the offloading feasibility tunnel $\mathcal{T}_B(\theta)$, the current tunnel has a lower floor, which can potentially reduce the transmission-energy consumption. Moreover, since $\mathcal{T}_B(\theta)$ can be regarded as a special case of the current tunnel $\bar{\mathcal{T}}_B(\theta)$ for which $\sum_{j=1}^{K-1} \theta L_j=U_{{\rm{bit}},K}$, the feasibility conditions for Problem P5 can be easily derived stated in the following lemma.
\begin{lemma}\label{Lem:FeaConP5}\emph{Problem P5 is feasible if and only if $0\le \theta \le \theta_{\max}$ where
\begin{equation}\label{Eq:MaxTheta}
\theta_{\max}=\min\l\{1, \min_k \l\{ \frac{U_{{\rm{bit}},K}-U_{{\rm{bit}},k}}{\sum_{j=k}^{K-1} L_j} \r\}
\r\}.
\end{equation}}
\end{lemma} 
Next, given Problem P5 is feasible, the lemma below states one important property of the defined effective offloading feasibility tunnel, proved by a similar method for Proposition~\ref{Pro:OptRegion}.

\begin{lemma}\label{Lem:BurstyOptRegion}\emph{Consider the helper has a large buffer and the user has bursty data arrivals for offloading. For the case of underutilization, the energy-efficient transmission policy can be derived by forming a shortest path in the effective offloading feasibility tunnel.
}
\end{lemma}
Thus, Problem P5 for the current case can be transformed to the one replacing the constraints with the effective offloading feasibility tunnel, and solved by the ``string-pulling" approach.

\subsection{Energy-Efficient Proportional Data Partitioning}

In this sub-section, the energy-efficient proportional data partitioning is transformed into the same form as the counterpart with one-shot data arrival and solved using a similar method.

First, consider the feasibility of Problem P6. It is feasible if and only if there exists one data-partitioning ratio, for which both the adaptive offloading and  local computing at the user are feasible. For each ratio, the former can be verified in the slave Problem P5 in the preceding sub-section and the latter is analyzed as follows. Similar to the effective offloading feasibility tunnel, given on the constraints of deadline and data causality for local computing, we define an effective local-computing feasibility tunnel as
\begin{align}\label{Eq:EffectiveLocTunnel}
& \text{(Effective Local-Computing Feasibility Tunnel)}\nn\\
&\bar{\mathcal{T}}_{\rm{B, loc}}(\theta)=\l\{ \boldsymbol{\ell}_{\rm{loc}}  ~\bigg|~\l[\frac{\hat{s}_k f}{C}-\bar{\Delta}_{\rm{loc}}(\theta)\r]^{+} \le \sum_{j=1}^k \ell_{\rm{loc},j}\le \sum_{j=1}^{k-1} (1-\theta)L_j,  \right.\nn\\ &   ~~~~~~~~~~~~~~~~~~~~~~~~~~~~~~\left.~\text{for}~ k=1,\cdots, \tilde{K}-1,~\text{and}~ \sum_{k=1}^{\tilde{K}} \ell_{{\rm{loc}},k} =\sum_{k=1}^{\tilde{K}-1} (1-\theta) L_k\r\}
\end{align}
where $\bar{\Delta}_{\rm{loc}}(\theta)=\frac{T f}{C}-\sum_{j=1}^{\tilde{K}-1} (1-\theta)L_j$. The local computing is feasible if and only if the tunnel ceiling is not below the tunnel floor. Combing the feasibility conditions for local computing and offloading yields the feasibility conditions for Problem P6 in the following lemma.
\begin{lemma}\label{Lem:FeaConP6}\emph{Problem P6 is feasible if and only if $\theta_{\min}\le \theta \le \theta_{\max}$ where
\begin{equation}
\theta_{\min}=\l[1- \min_k \l\{ \frac{f(T-\hat{s}_k)/C}{\sum_{j=k}^{\tilde{K}-1} L_j} 
%, \frac{Tf}{C \sum_{j=1}^{\tilde{K}-1} L_j}
\r\}\r]^+ 
\end{equation}
and $\theta_{\max}$ is defined in \eqref{Eq:MaxTheta}.}
\end{lemma}
Using Lemma~\ref{Lem:FeaConP6}, Problem P6 can be transformed as:
\begin{equation*}({\bf P7})\qquad
\begin{aligned}
\min_ {\theta}   \quad& \l[\sum_{k=1}^{\tilde{K}-1} (1-\theta) L_k\r] C P_{\rm{cyc}}+\hat{E}_{\rm{off}}(\theta)
\qquad \rm{s.t.}~~
& \theta_{\min}\le\theta\le \theta_{\max}.
\end{aligned}
\end{equation*}
 Problem P7 has a similar form as that of Problem P2. Using the similar approach, Problem P7 can be proved to be a convex problem and the optimal data-partitioning ratio can be computed using the sub-gradient method. The details are omitted for brevity.

\section{Simulation Results}\label{Sec:Simu}
%In this section, the performance of the proposed co-computing design is evaluated by simulation. 
The simulation parameters are set as follows unless specified otherwise. First, the computation deadline is set as $T = 0.1$ s.  For local computing, the CPU frequency is $f=1$ GHz. The required number of CPU cycles per bit is $C=500$ cycle/bit and each CPU cycle consumes energy $P_{\text{cyc}}=10^{-10}$ J with $\gamma\!=\!10^{-28}${\color{black}\cite{chen2015efficient,you2016energy}}. For offloading, {\color{black}{we assume that the signal attenuation from the user to the helper is $60$ dB corresponding to an equal distance of $10$ meter, and the channel $h$ is generated from Rayleigh fading \cite{ju2014throughput}.}} 
%Let $\delta$ denote the distance-dependent path loss, modeled by $\delta=\delta_0 (d/d_0)^{-\alpha}$ \cite{}, where $d$ is the distance from the user to the helper, $\delta_0=10^{-3}$, $d_0=1$ m is the reference distance, and $\alpha$ is the path loss exponent set to be $3$ \cite{}. The small-scale fading is  set as $h=10^{-3}$,
Moreover,  the bandwidth $B=1$ MHz and the variance of complex-white-Gaussian-channel noise  $N_0=-70$ dBm. Next, for the helper, its CPU frequency is $f_h=5$ GHz. The helper-CPU state alternates between idle and busy. Both the  idle and busy intervals follow independent exponential distributions where the expected busy interval fixed as $0.02$ s and the expected idling interval being a variable. 
%Moreover, it reserves  $Q=1$ Kb buffer for the co-computing.

\subsection{One-Shot Data Arrival}

Consider the case where the user has one-shot input data arrival and the helper has a large buffer.
  We evaluate the performance of \emph{computing probability} and user's energy consumption. Specifically, computing probability is defined as the probability that the user finishes the given computation load via simultaneous offloading and local computing. For comparison, a \emph{benchmark policy} is considered, for which the P2P transmission rate follows the helper's CPU-idling profile and the data partitioning is optimized using the sub-gradient algorithm. 

\begin{figure}[t!]
\centering
\subfigure[Computing probability.]{\label{Fig:Simu_Successful}
\includegraphics[width=8cm]{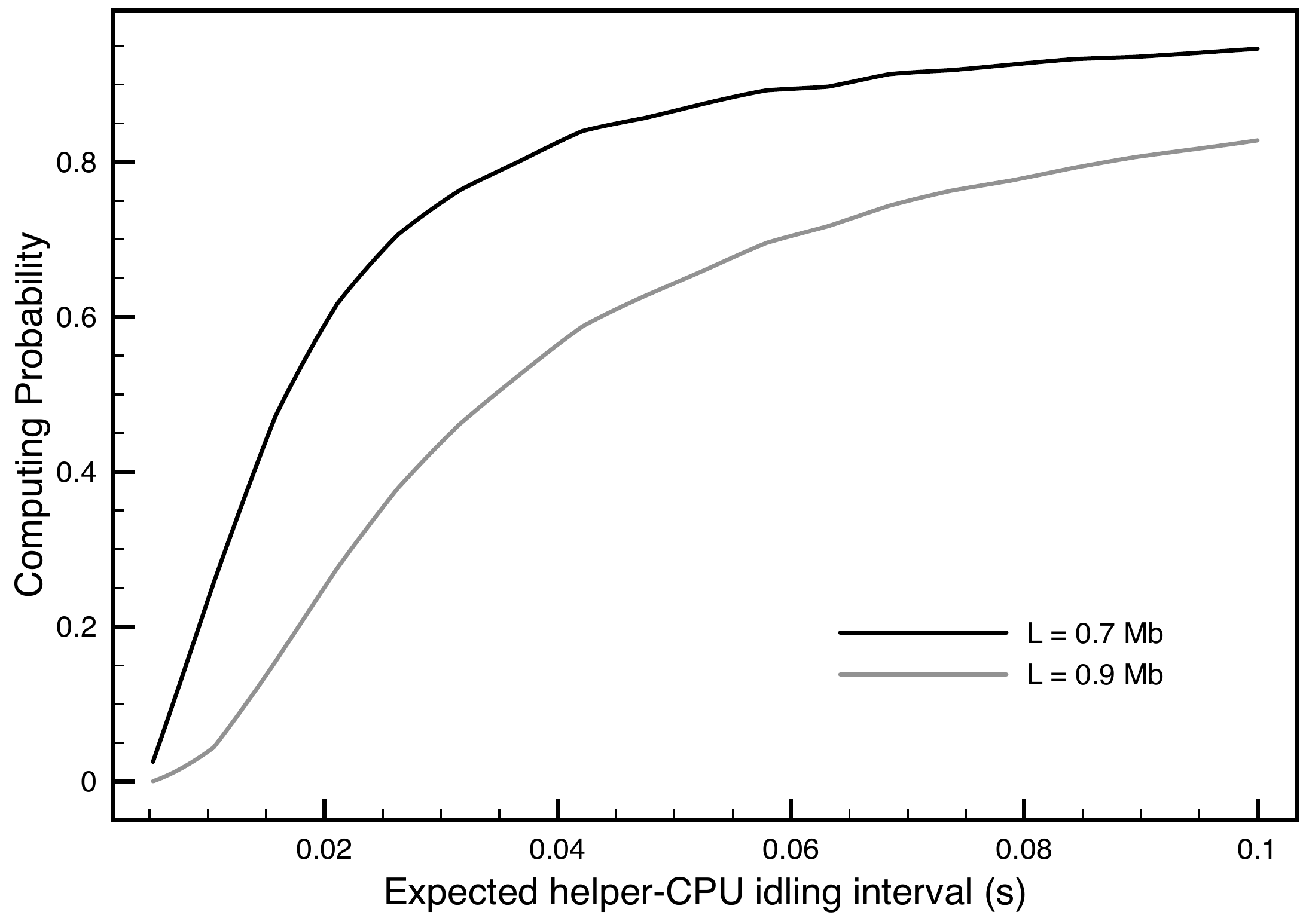}}
\subfigure[User's energy consumption.]{\label{Fig:Simu_Egy}
\includegraphics[width=8cm]{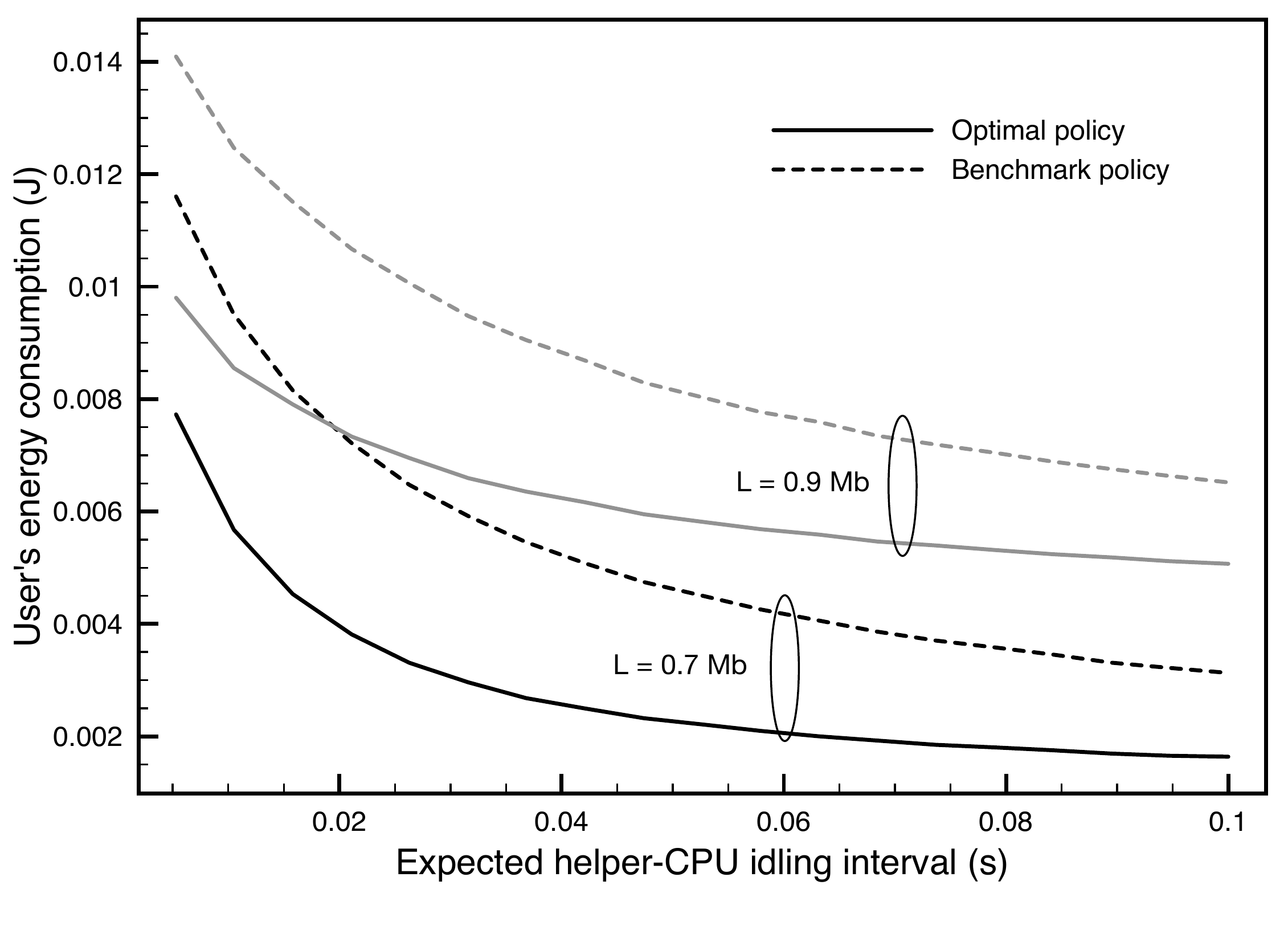}}
\caption{\color{black}{Effects of helper-CPU idling interval on the computing probability and user's energy consumption for the case of one-shot data arrival and a large buffer at the helper.}}
\end{figure}

Fig.~\ref{Fig:Simu_Successful} shows the curves of computing probability versus the expected helper-CPU idling interval. One can observe that the computing probability increases when the user has  the decreasing computing load $L$ or the increasing idling interval.  Moreover, computing probability grows at a higher rate when the helper has a relatively small expected CPU idling interval. 

The curves of the user's energy consumption versus the expected helper-CPU idling intervals  are plotted in Fig.~\ref{Fig:Simu_Egy}. Several observations are made as follows. First, the energy consumption is monotone-decreasing with the growing of helper-CPU idling interval since it allows the user to reduce the transmission rate for reducing transmission-energy consumption. However, the energy consumption saturates when the expected helper-CPU idling interval is large. Next, observe that  the optimal policy achieves substantially higher energy savings compared with the benchmark policy since the former exploits the helper-CPU busy intervals for P2P transmission. 

%Last, the effects of buffer size on the user's energy consumption are analyzed by simulation as shown in the extended version \cite{you2017energy}.

\begin{figure}[t!]
%\vspace{20pt}
\begin{center}
\includegraphics[width=8cm]{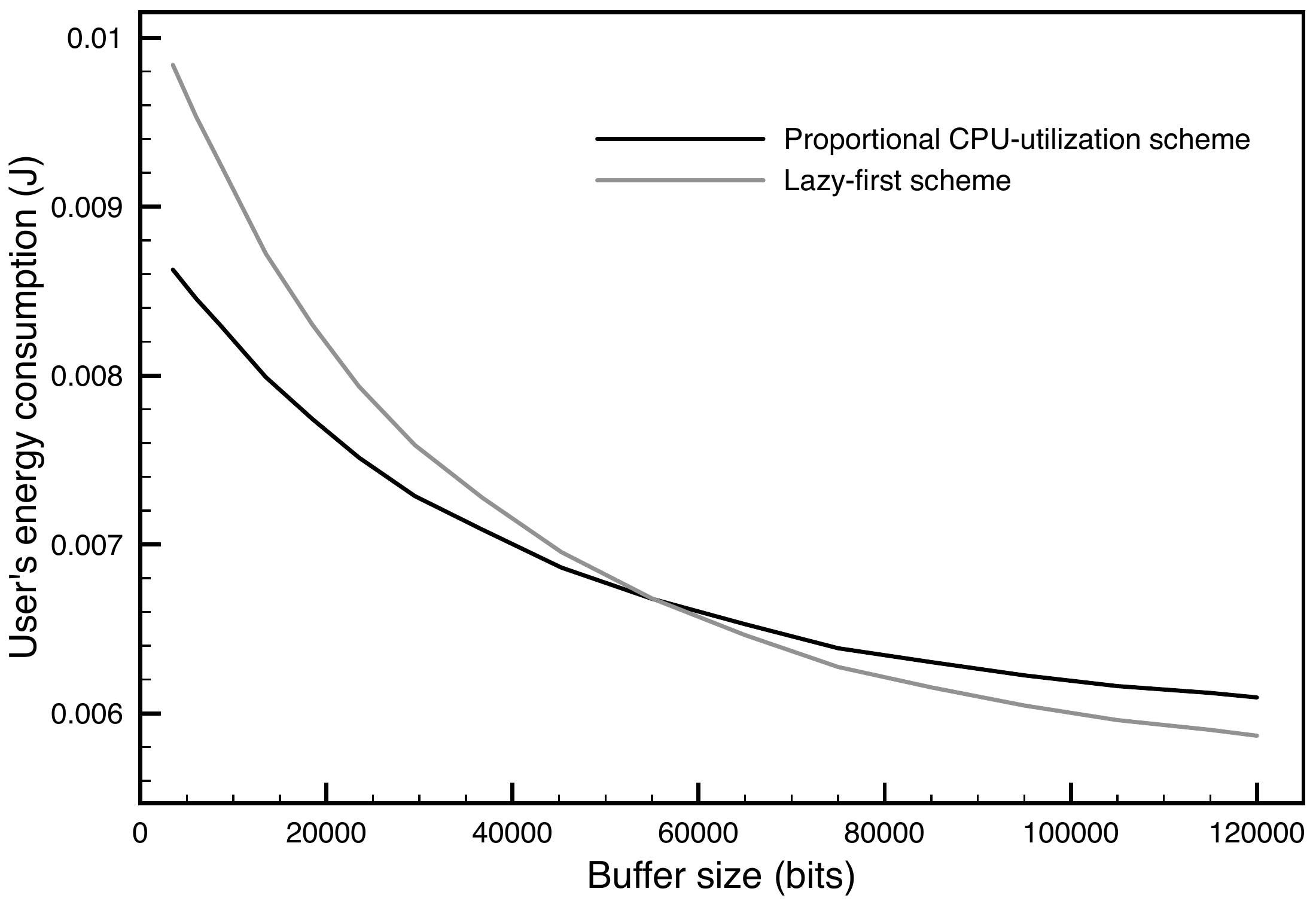}
\caption{\color{black}{Effects of buffer size on the user's energy computation.}}
\label{Fig:Simu_Egy_Buffer}
\end{center}
\end{figure}

Last, the effects of buffer size on the user's energy consumption are shown in Fig.~\ref{Fig:Simu_Egy_Buffer}. Consider one baseline \emph{lazy-first} scheme which postpones the CPU co-computing in the early idle epochs and then fully utilizes the helper's CPU resource in the later epochs. The computation load is set as $L=0.7$ Mb. One can observe that with the grow of the buffer size, user's energy consumption firstly decreases owing to the buffer gain and then saturates when the buffer size is large. Next, compared with the lazy-first scheme, the proposed scheme of proportional CPU-utilization contributes to less user's energy consumption when the buffer size is small but has more energy consumption when the buffer exceeds a threshold ({\color{black}{about $0.55$ Mb}}). The reason is that for the former case, the offloading policy tends to follow the helper's CPU profile, and the proportional CPU-utilization scheme can distribute the buffer gain to all idle epochs and thereby lead to less variation on the offloading rates. While when the buffer is sufficiently large, the lazy-first scheme is the optimal policy as shown in Section~\ref{Sec:UnderutilizationEgy}. This observation is coincident with Remark~\ref{Rem:BufferGain}. Other observations are similar to those from Fig.~\ref{Fig:Simu_Egy}.

\vspace{-5pt}
 \subsection{Bursty Data Arrivals}
Consider the case where user has bursty data arrivals. Specifically, the data inter-arrival interval follows the exponential distribution  and for each arrival, the data size is uniformly distributed. The expected helper-CPU idling interval is set as $0.02$ s. Consider a benchmark policy for performance comparison, for which the adaptive offloading follows the curve of the floor of effective offloading feasibility tunnel and the proportional data partitioning is optimized using the sub-gradient algorithm.

\begin{figure}[t!]
\centering
\subfigure[Computing probability.]{\label{Fig:Simu_Successful_Bursty}
\includegraphics[width=8cm]{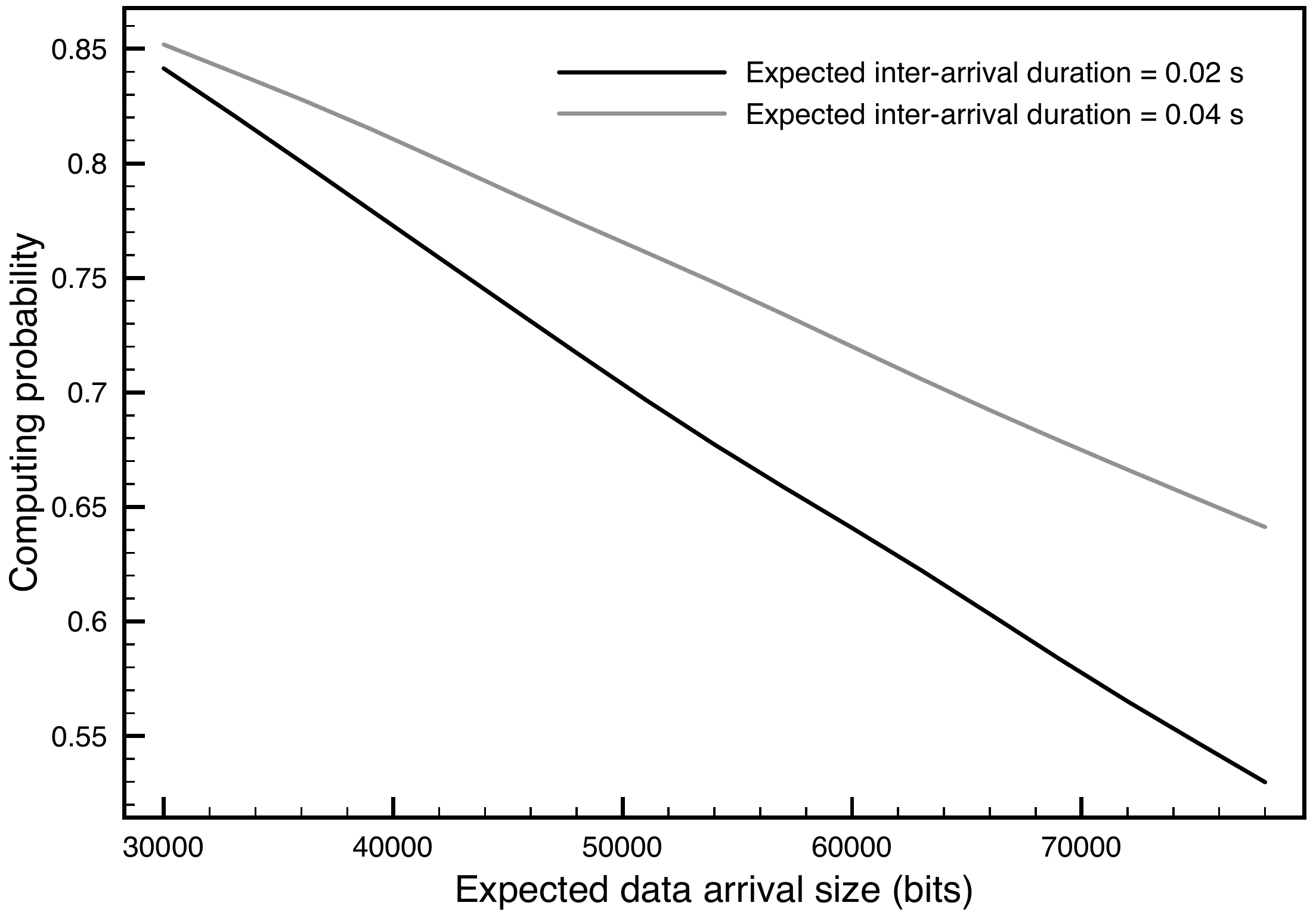}}
\subfigure[User's energy consumption.]{\label{Fig:Simu_Egy_Bursty}
\includegraphics[width=8cm]{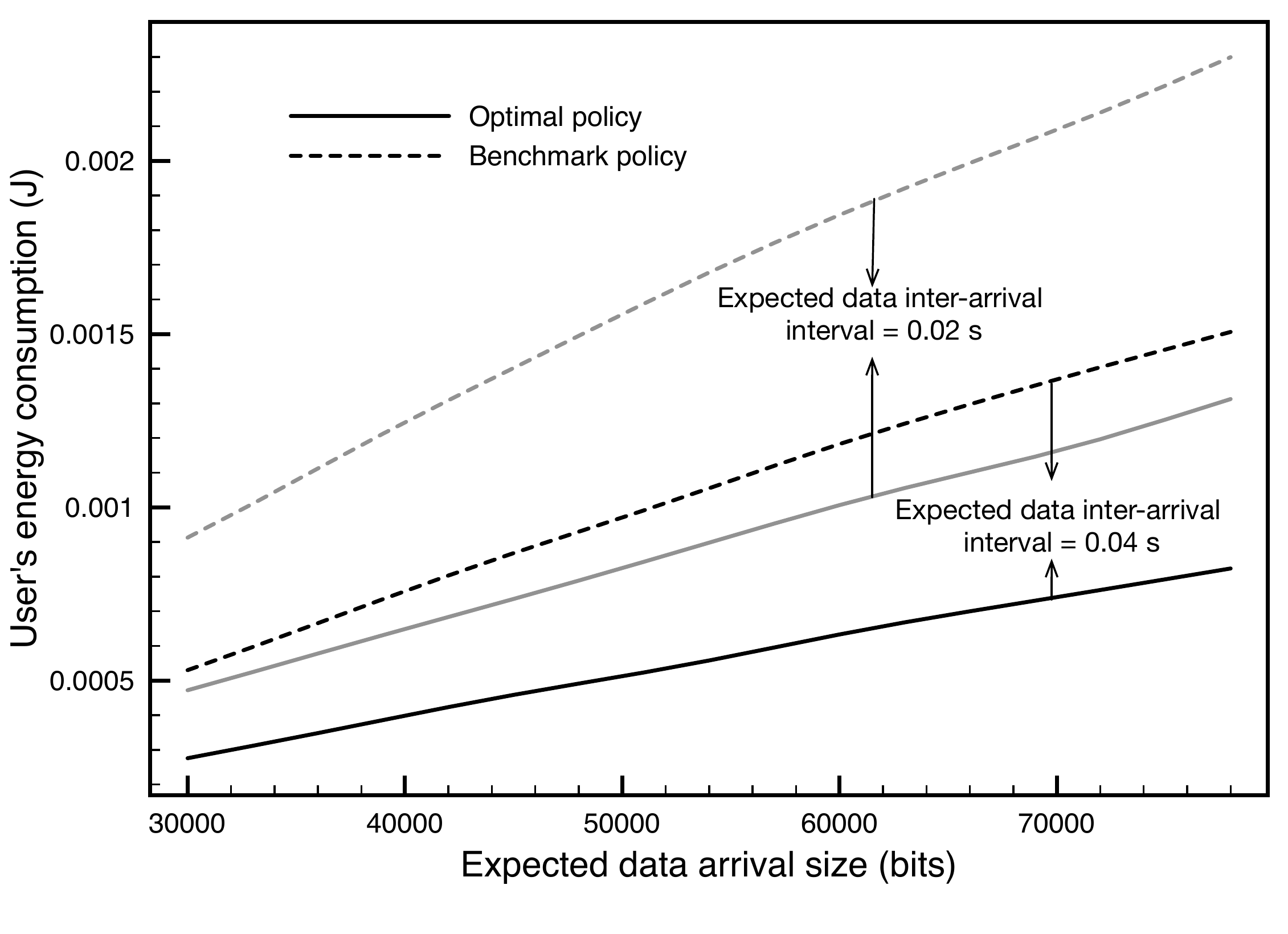}}
\caption{\color{black}{Effects of user's expected data arrival size on the computing probability and user energy consumption for the case of bursty data arrivals.}}
\end{figure}
Fig.~\ref{Fig:Simu_Successful_Bursty} depicts the curves of computing probability versus the user's expected data arrival size under different expected data inter-arrival intervals. It is interesting to observe that the computing probability decreases \emph{linearly} with the user's expected data arrival size. Moreover, the decreasing rate is higher when the user has more frequent data arrivals resulted from a shorter expected data inter-arrival duration. 

The curves of  user's average energy consumption versus the expected data arrival size are shown in Fig.~\ref{Fig:Simu_Egy_Bursty}.  One can observe that the user's energy consumption is almost \emph{linearly} increasing with the grow of expected data arrival size. Moreover, the energy consumption grows more sharply when the user has more frequent data arrivals.  Last, the optimal policy achieves higher energy savings compared with the benchmark policy, especially when the user has a large data arrival rate. 

{\color{black}{\section{Conclusion}
In this paper, we have presented a new design for mobile cooperative computing that enables a user to exploits NC-CSI shared by a cooperative helper for  fully utilizing random computation resources at the helper by offloading with minimum energy consumption. The designed optimal policies for controlling offloading have been formulated as constrained optimization  problems and solved using convex optimization theory. Thereby, we have re-discovered the optimal ``string-pulling" structure in the polices that also lies in those for controlling  transmissions for energy-harvesting systems. This work opens a new direction for mobile cooperative computing, namely applying computation prediction to enable scavenging of random computation resources at edge devices. Along this direction, there lie many promising opportunities. In particular,  the current design for a single user-helper pair can be extended to  complex co-computing networks, addressing design issues such as applying computation prediction to facilitate joint radio-and-computation resource allocation and helper probing.}}

\appendix
 \subsection{Proof of Proposition~\ref{Pro:OptRegion}}\label{App:OptRegion}
Since $\sum_{j=1}^k d_j(\ell_j) \le \sum_{j=1}^k \ell_j\le \ell$ and $[U_{{{\rm{bit}}},k}-\Delta(\ell)]^{+} \le \sum_{j=1}^k d_j(\ell_j)\le U_{{{{\rm{bit}}}},k}$. The optimal offloading policy solving Problem P3 can be derived by: i) For each offloading feasibility tunnel with floor $\sum_{j=1}^k d_j(\ell_j) \le \ell_j$, compute its shortest path as a candidate for which  the optimal offloaded policy also satisfies \eqref{Eq:BEvolve}; ii) Selecting the shortest path over all candidates. This policy  is also the shortest path over the offloading feasibility tunnel $\bar{\mathcal{T}}(\ell)$. The corresponding tunnel floor can be computed from the optimal policy $\boldsymbol{\ell}^*$ using \eqref{Eq:BEvolve}, completing the proof.
 \hfill $\blacksquare$
 
 \subsection{Proof of Proposition~\ref{Prop:ProCPUUtil}}\label{App:ProCPUUtil}
Let $\tilde{f}_{h,k}$ denote the number of allocated CPU cycles per second in epoch $k$ and $A$ the set of idle epoch indexes. It is equivalent to prove that $\tilde{f}_{h,k}=\tilde{f}_h, \forall k\in A$. First, when the buffer size $Q\to 0$, the offloaded data should be immediately computed, i.e., $d_k(\ell_k)=\dfrac{a_k \tau_k \tilde{f}_h}{C}=\ell_k\le \dfrac{a_k \tau_k f_h}{C}$. Thus, the shortest path is the curve of offloaded bits, whose length can be given as $\sum_{k=1}^K \sqrt{d_j(\ell_k)^2+\tau_k^2}$. In particular, the length for the busy epoch is fixed as $\tau_k$ since $d_j(\ell_k)=0$. Thus, given the fixed computed bits, the problem for the shortest path can be formulated as
\begin{equation*}({\bf P8})\qquad
\begin{aligned}
\min_ {\ell_k}   \quad& \sum_{k\in A} \sqrt{d_j(\ell_k)^2+\tau_k^2}
\qquad \rm{s.t.} 
& \sum_{k\in A} d_j(\ell_k)=\ell.
\end{aligned}
\end{equation*}
It is easy to derive that the optimal solution for Problem P8 satisfies: $\dfrac{d_k(\ell_k)}{\ell}=\dfrac{\tau_k}{\sum_{k\in A} \tau_k}$, $\forall k\in A$.
Therefore, for each $k\in A$, it has
$$\tilde{f}_{h,k}=\frac{d_k(\ell_k)C}{\tau_k}=\frac{\ell C}{\sum_{k\in A} \tau_k}=f_h \ell \frac{C}{f_h \sum_{k\in A} \tau_k}=f_h \frac{\ell}{U_{{\rm{bit}},K}}=\tilde{f}_{h},$$ completing the proof.
 \hfill $\blacksquare$

\subsection{Proof of Lemma~\ref{Lem:PropRegi}}\label{App:PropRegi} 
We first introduce a lemma below to facilitate the proof which can be proved easily.
\begin{lemma}\label{Lem:MaxMin}\emph{Given constants $a, b, c$ and $d$, it has $\max\{a,b\}+\max\{c,d\}\ge \max\{a+c,b+d\}.$
}
\end{lemma} 
Then, it is equivalent to prove that the construction of a policy $\lambda \boldsymbol{\ell}^{(1)} + (1-\lambda) \boldsymbol{\ell}^{(2)}$ satisfies the constraints in the offloading feasibility region $\bar{\mathcal{T}}(\lambda \ell^{(1)} + (1-\lambda) \ell^{(2)})$, as proved below.

First, since $\boldsymbol{\ell}^{(1)} \in \bar{\mathcal{T}}(\ell^{(1)})$ and $\boldsymbol{\ell}^{(2)} \in \bar{\mathcal{T}}(\ell^{(2)})$, it has $[U_{{\rm{bit}},k}-\Delta(\ell^{(1)})]^{+}\le \sum_{j=1}^k \ell_j^{(1)} \le \ell^{(1)},~[U_{{\rm{bit}},k}-\Delta(\ell^{(2)})]^{+}\le \sum_{j=1}^k \ell_j^{(2)} \le \ell^{(2)}$, for $k=1,\cdots, K$; $\sum_{j=1}^{K} \ell_j^{(1)}=\ell^{(1)}$ and $\sum_{j=1}^{K} \ell_j^{(2)}=\ell^{(2)}$.
Next, for the constructed policy  $\lambda \boldsymbol{\ell}^{(1)} + (1-\lambda) \boldsymbol{\ell}^{(2)}$, we have:  $\sum_{j=1}^k [\lambda \ell_j^{(1)}+(1-\lambda)\ell_j^{(2)}] =\lambda \sum_{j=1}^k \ell_j^{(1)}+(1-\lambda) \sum_{j=1}^k \ell_j^{(2)}$. 
Combing the above results and Lemma~\ref{Lem:MaxMin} yields
\begin{align*}
\sum_{j=1}^k [\lambda \ell_j^{(1)}+(1-\lambda)\ell_j^{(2)}] &\ge \lambda [U_{{\rm{bit}},k}-\Delta(\ell^{(1)})]^{+}+(1-\lambda)[U_{{\rm{bit}},k}-\Delta(\ell^{(2)})]^{+}
 \nn\\
 &
 \ge \{U_{{\rm{bit}},k}-\Delta(\lambda \ell^{(1)}+(1-\lambda)\ell^{(2)})\}^{+},  ~~&k=1,\cdots, K
 \\
  &\!\!\!\!\!\!\!\!\!\!\!\!\!\!\!\!\!\!\!\!\!\!\!\!\!\!\!\!\!\!\!\!\!\!\!\!\!\!\!\!\!\!\!\!\!\!\!\!\!\!\!\!\!\!\!\!\!\!\!  \sum_{j=1}^k [\lambda \ell_j^{(1)}+(1-\lambda)\ell_j^{(2)}] \le \lambda\ell^{(1)}+(1-\lambda) \ell^{(2)},~~&k=1,\cdots, K
\end{align*}
and $\sum_{j=1}^K \lambda \ell_j^{(1)}+(1-\lambda)\ell_j^{(2)} =\lambda \ell^{(1)}+(1-\lambda)\ell^{(2)}$.
Thus, the policy $\lambda \boldsymbol{\ell}^{(1)} + (1-\lambda) \boldsymbol{\ell}^{(2)}$ satisfies all the constraints, completing the proof.
\hfill $\blacksquare$

\subsection{Proof of Lemma~\ref{Lem:MasConvTime}}\label{App:MasConvTime} 
Let $\boldsymbol{\ell}^*$, ${\boldsymbol{\ell}^{(1)}}^*$ and ${\boldsymbol{\ell}^{(2)}}^*$ denote the optimal offloading policies for the offloaded data size $\lambda \ell^{(1)}+(1-\lambda) \ell^{(2)}$, $\ell^{(1)}$ and $\ell^{(2)}$, respectively. From the definition of $E_{\rm{off}}(\ell)$, we have the following:
\begin{align*}
E_{\rm{off}}( \lambda \ell^{(1)} + (1-\lambda) \ell^{(2)})=&\sum\limits_{k=1}^{K} \dfrac{\tau_k}{h^2} f\!\l(\dfrac{\ell_k^*}{\tau_k}\r)
\overset{(a)}{\le}\sum\limits_{k=1}^{K} \dfrac{\tau_k}{h^2} f\!\l(\dfrac{\lambda {\ell_k^{(1)}}^*+(1-\lambda){\ell_k^{(2)}}^* }{\tau_k}\r)\\
\overset{(b)}{\le} & \sum\limits_{k=1}^{K} \dfrac{\tau_k}{h^2}\l[ \lambda  f\!\l(\dfrac{{\ell_k^{(1)}}^*}{\tau_k}\r)+(1-\lambda)  f\!\l(\dfrac{{\ell_k^{(2)}}^*}{\tau_k}\r)\r]\\
=& \lambda E_{\rm{off}}( \ell^{(1)})+(1-\lambda)E_{\rm{off}}(\ell^{(2)})
\end{align*}
where $(a)$ is due to that the constructed policy $\lambda \boldsymbol{\ell}^{(1)} + (1-\lambda) \boldsymbol{\ell}^{(2)}$ is feasible given offloaded data size $\lambda \ell^{(1)} +(1-\lambda) \ell^{(2)}$ as shown in Lemma~\ref{Lem:PropRegi} but can be sub-optimal, and $(b)$ is due to the convexity property of the function $f(x)$, leading to the desired result.
 \hfill $\blacksquare$

%  \bibliographystyle{ieeetr}
%\bibliography{BibDesk_File}

\end{document}